\documentclass[11pt]{article}

\usepackage[T1]{fontenc}
\usepackage[utf8]{inputenc}
\usepackage{lmodern}
\usepackage{microtype}
\usepackage[a4paper,margin=30mm]{geometry}

\usepackage{amsmath,amssymb,amsthm}
\usepackage{booktabs}
\usepackage{array}
\usepackage{graphicx}
\usepackage{tikz}
\usetikzlibrary{arrows.meta}
\usepackage{natbib}
\usepackage{url}
\usepackage{fancyvrb}
\usepackage[hidelinks]{hyperref}

\newcolumntype{L}[1]{>{\raggedright\arraybackslash}p{#1}}

\newtheorem{proposition}{Proposition}
\newtheorem{lemma}[proposition]{Lemma}
\newtheorem{corollary}[proposition]{Corollary}

\newcommand{\pkg}[1]{\textsf{#1}}
\newcommand{\proglang}[1]{\textsf{#1}}
\newcommand{\code}[1]{\texttt{#1}}
\newcommand{\class}[1]{\code{#1}}
\newcommand{\fct}[1]{\code{#1()}}
\DefineVerbatimEnvironment{CodeInput}{Verbatim}{fontsize=\small,fontshape=sl}
\DefineVerbatimEnvironment{CodeOutput}{Verbatim}{fontsize=\small}
\newenvironment{CodeChunk}{\par\smallskip}{\par\smallskip}

\newcommand{\Ften} {\fontsize{10}{11}\selectfont  }

\usepackage{fancyhdr}
\title{\pkg{GraphInfer}: Statistical Inference for Decomposable Graphical Models in \proglang{R}}

\author{
  Rodrigo Labouriau
  \footnote{Department of Mathematics, Aarhus University and RLstatLab. \\
  E-mail:
  \texttt{rodrigo.labouriau@rlstatlab.com}}
}

\date{}

\begin{document}

\maketitle

\begin{abstract}
\noindent
Decomposable graphical models combine closed-form maximum likelihood estimation
with local model comparison, which makes structure inference practical with many
variables. The \proglang{R} package \pkg{GraphInfer} provides one likelihood
framework for decomposable models of continuous, discrete and mixed data, the last
as conditional Gaussian models. Structure inference ranges from exact minimum
forests and stepwise search to stochastic global search. Nested models are compared
by likelihood-ratio tests whose null distributions are exact in finite samples:
conditionally on the discrete data for edges with a continuous endpoint, and
conditionally on the discrete margins of the smaller model for edges between
discrete variables.
Missing values are handled by EM, by structural EM, which reuses the complete-data
searches through a weighted pseudo-sample of expected sufficient statistics, and by
multiple imputation; selected models are fit to the observed data likelihood.
The package also treats the graphical models induced by irreversible event
accumulation on trees: a search over the compatible graphs whose tree moves overcome
an obstruction to single-edge moves, fitting of the full accumulation model with
measurement noise and bootstrap tests of the accumulation hypothesis.
\pkg{GraphInfer} is an independent, pure \proglang{R} successor to the archived
package \pkg{gRapHD}. The implementation is validated against independent
references and simulation studies inform its defaults.
\end{abstract}

\medskip
\noindent
\textbf{Keywords:}
graphical models; decomposable models; conditional Gaussian distributions;
structure inference; likelihood inference; missing data; accumulation trees;
\proglang{R}.

%\medskip
%{\setcounter{tocdepth}{1}\tableofcontents}
%\bigskip

\newpage
\Ften
\tableofcontents
\newpage
\normalsize

\section{Introduction}
\label{sec:intro}

Undirected graphical models describe the conditional independences among a set of
variables through a graph \citep{Lauritzen:1996,Edwards:2000,Whittaker:1990}. Among
them, decomposable models have a special computational position: their likelihood
factorises over the cliques and separators of the graph, their maximum-likelihood
estimates are available in closed form, and the effect of adding or removing one
edge can be obtained from a few low-dimensional marginal models. These properties
extend to mixed data through conditional Gaussian distributions, under the stronger
requirement of strong decomposability
\citep{Lauritzen+Wermuth:1989,Frydenberg+Lauritzen:1989}. They are what make a
search over graphs feasible when the number of variables is large.

The \proglang{R} package \pkg{GraphInfer} supports the whole of an analysis with
such models: a first overview of the dependence structure, the search for a graph,
the fitting and testing of selected models, and the treatment of incomplete
observations, for continuous, discrete and mixed variables. Its contribution can be
organised around four themes.
\begin{enumerate}
  \item \emph{A common inferential framework.} Decomposable Gaussian, discrete and
  conditional Gaussian models are handled by one likelihood calculation, with one
  convention for model dimension, shared by all fitting, testing and search
  procedures (Sections~\ref{sec:framework} and~\ref{sec:inference}). This includes
  likelihood-ratio tests that are exact in finite samples, for edges with a continuous
  endpoint and, conditionally on the discrete margins, for edges between discrete
  variables.
  \item \emph{Structure inference at several levels of effort}, from minimum forests,
  where the optimum is obtained exactly, through local stepwise procedures to
  non-greedy global searches (Section~\ref{sec:structure}).
  \item \emph{Inference with missing values}, by EM for a given graph, by structural
  EM for an unknown graph, and by imputation (Section~\ref{sec:missing}).
  \item \emph{Models with a substantive origin.} The graphical models induced by
  irreversible event accumulation on trees \citep{Labouriau:AccumulationTrees} show
  how the framework extends to a restricted, scientifically motivated class of graphs
  (Section~\ref{sec:acctrees}).
\end{enumerate}

We distinguish three levels throughout: the statistical model, decomposable
Gaussian, discrete or conditional Gaussian distributions
(Section~\ref{sec:framework}); the inferential procedures, maximum likelihood, model
comparison, structure selection, EM and exact tests
(Sections~\ref{sec:structure} to~\ref{sec:acctrees}); and their computational
implementation, whose details are collected in Appendix~\ref{app:structure}. Much of
the package rests on classical results: the closed-form estimates and the
clique--separator decomposition of decomposable models, the minimum-forest
construction \citep{Chow+Liu:1968}, the exact Beta distributions of Gaussian
likelihood-ratio tests \citep{Lauritzen:1996}, EM for graphical models
\citep{Lauritzen:1995}, structural EM \citep{Friedman:1998} and multiple imputation
\citep{Rubin:1987}. Other parts are new implementations of established methods: an
accelerated stepwise engine, iterated local search over (strongly) decomposable
graphs, a row-wise EM for large discrete tables, the extended BIC for conditional
Gaussian models, and likelihood-ratio tests whose degrees of freedom compare both
models on one support. Exact likelihood-ratio tests of a single edge in mixed models are classical
\citep{Frydenberg+Lauritzen:1989,Lauritzen:1996,Edwards:2000,Tur+Castelo:2011,
Tur+Roverato+Castelo:2014}, and \citet[Proposition~6.27]{Lauritzen:1996} states, with
an outline of the proof, that along a sequence of single-edge removals the
likelihood-ratio factors of edges with a continuous endpoint are independent
conditionally on the counts. We give a complete proof, with the parameters in terms of
occupied cells, and implement the resulting exact tests for nested mixed models
differing by several edges, homogeneous or heterogeneous (Section~\ref{sec:exact},
Appendix~\ref{app:exactproofs}). To our knowledge, three results are new: the weighted pseudo-sample by which structural EM
reuses the complete-data searches for conditional Gaussian models
(Section~\ref{sec:sem}); an obstruction that prevents single-edge moves from
reaching some accumulation-tree compatible graphs, and the connectivity of the tree
moves that overcome it (Section~\ref{sec:acctree-search}); and, for accumulation on a
tree, a likelihood in which each unit follows a single lineage that stops at a node, with
event-specific measurement noise, fitted by EM and compared at two levels with
decomposable graphical models (Section~\ref{sec:acctree-full}). Accumulation under
observation error has a long history in models of tumour progression
\citep{Szabo+Boucher:2002,Gerstung+others:2009}; Section~\ref{sec:acctree-full} sets out
how the formulation here is different.

The minimum-forest construction and its extension to penalised criteria and mixed
data \citep{Edwards+Abreu+Labouriau:2010} were made available, together with stepwise
search, in the \proglang{R} package \pkg{gRapHD}
\citep{Abreu+Edwards+Labouriau:2010}, since archived on CRAN. \pkg{GraphInfer} is its
independent successor: it was written from the published statistical theory, contains
no code from \pkg{gRapHD}, and retains the calling conventions of \fct{minForest} and
\fct{stepw} so that existing analyses can be migrated. It is a readable reference
implementation in pure \proglang{R}: it contains no compiled code, depends only on the
base packages \pkg{stats} and \pkg{utils}, and uses \pkg{igraph} only optionally, for
plotting. Its speed comes from the local structure of decomposable models rather than
from a compiled language. A planned companion package, \pkg{GraphInferComplex}, is
intended to speed up the identified computational kernels for the larger problems.

The closest related software is \pkg{gRbase} \citep{Dethlefsen+Hojsgaard:2005} and
\pkg{gRim} \citep{Hojsgaard+Edwards+Lauritzen:2012}. \pkg{gRim} covers discrete,
Gaussian and mixed interaction models, and offers forward and backward stepwise
selection among decomposable or unrestricted models. Its model class is therefore
wider. In the version we examined (0.3.4) it does not provide forest or global
search, or estimation from incomplete observations, and it is in these respects that
\pkg{GraphInfer} differs in inferential scope. The \pkg{Bioconductor} package
\pkg{qpgraph} implements exact likelihood-ratio tests of conditional independence for
homogeneous mixed data within its limited-order search \citep{Tur+Roverato+Castelo:2014}.
Penalised estimators such as the
graphical lasso \citep{Friedman+Hastie+Tibshirani:2008} address a different problem:
they estimate a sparse Gaussian concentration matrix without restricting the graph to
be decomposable.

Section~\ref{sec:overview} gives an overview of the package. Section~\ref{sec:framework}
The model class is defined, the likelihood is given and the selection criteria are stated.
Section~\ref{sec:structure} describes structure inference, Section~\ref{sec:inference}
the fitting and testing of a given model, Section~\ref{sec:missing} missing values, and
Section~\ref{sec:acctrees} accumulation trees. Section~\ref{sec:using} illustrates the
package on five data sets, and Section~\ref{sec:numerical} reports numerical studies of
its statistical and computational behaviour. Section~\ref{sec:discussion} concludes.
The appendices describe the architecture and implementation, the validation, and the
proofs for the exact tests in mixed models.

\section{Overview of the package}
\label{sec:overview}

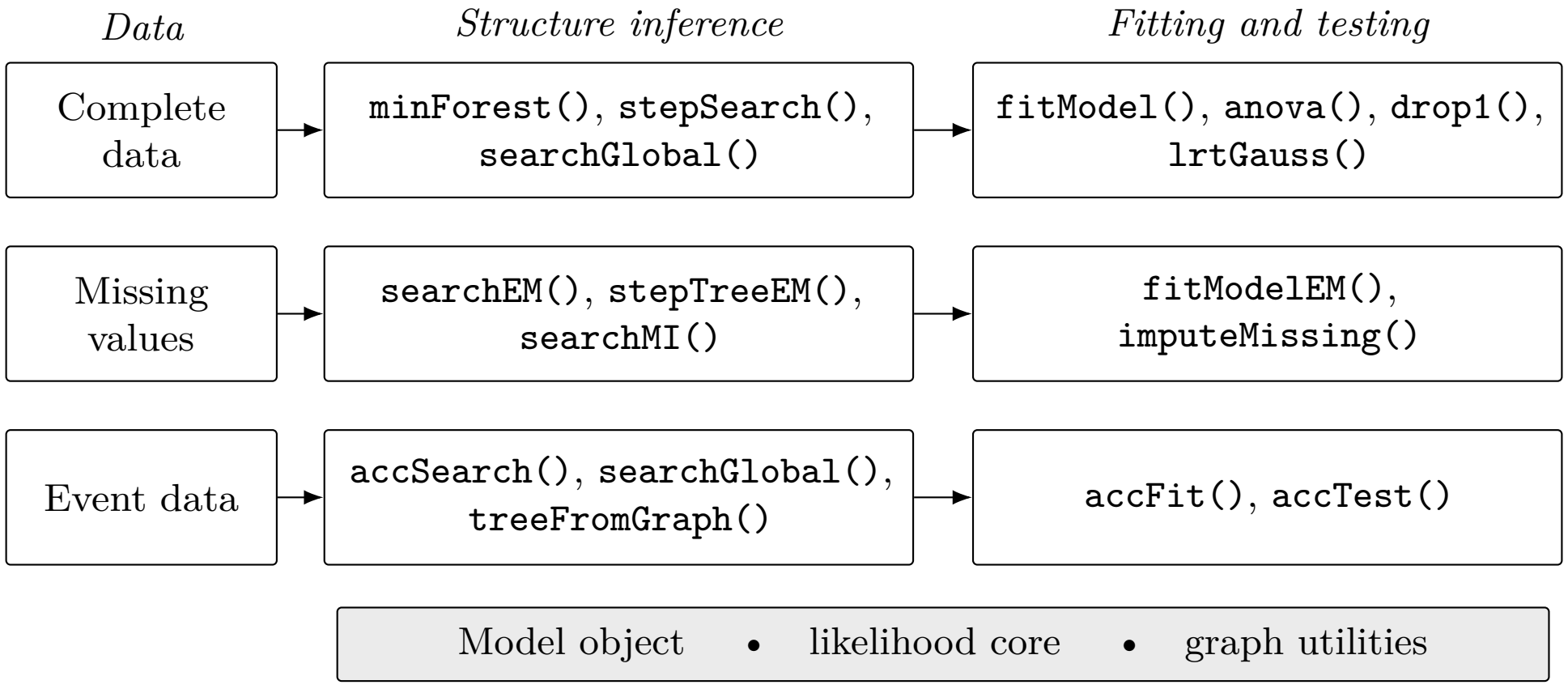
\begin{figure}[t]
\centering
\footnotesize
\begin{tikzpicture}[
  box/.style={draw, rounded corners=1pt, align=center, inner sep=3pt,
              minimum height=11mm, text width=#1},
  box/.default=38mm,
  bar/.style={draw, fill=black!8, rounded corners=1pt, align=center, inner sep=3pt,
              minimum height=6mm},
  arr/.style={-{Latex[length=1.6mm]}}]
\node[box=20mm] (d1) at (0,0) {Complete data};
\node[box=20mm] (d2) at (0,-15mm) {Missing values};
\node[box=20mm] (d3) at (0,-30mm) {Event data};
\node[box=46mm] (s1) at (39mm,0)
  {\fct{minForest}, \fct{stepSearch},\\ \fct{searchGlobal}};
\node[box=46mm] (s2) at (39mm,-15mm)
  {\fct{searchEM}, \fct{stepTreeEM},\\ \fct{searchMI}};
\node[box=46mm] (s3) at (39mm,-30mm)
  {\fct{accSearch}, \fct{searchGlobal},\\ \fct{treeFromGraph}};
\node[box=46mm] (f1) at (92mm,0)
  {\fct{fitModel}, \fct{anova}, \fct{drop1},\\ \fct{lrtGauss}};
\node[box=46mm] (f2) at (92mm,-15mm)
  {\fct{fitModelEM}, \fct{imputeMissing}};
\node[box=46mm] (f3) at (92mm,-30mm)
  {\fct{accFit}, \fct{accTest}};
\foreach \i in {1,2,3} {
  \draw[arr] (d\i) -- (s\i);
  \draw[arr] (s\i) -- (f\i);
}
\node[font=\footnotesize\itshape] at (0,8.5mm) {Data};
\node[font=\footnotesize\itshape] at (39mm,8.5mm) {Structure inference};
\node[font=\footnotesize\itshape] at (92mm,8.5mm) {Fitting and testing};
\node[bar, minimum width=99mm] at (65.5mm,-42mm)
  {Model object \quad\textbullet\quad likelihood core \quad\textbullet\quad graph utilities};
\end{tikzpicture}
\caption{Main routes through the package. For event data \fct{searchGlobal} runs with
\code{compatible = TRUE}. The searches of all three rows score graphs through the same
likelihood core.}
\label{fig:workflow}
\end{figure}

Figure~\ref{fig:workflow} shows the main routes through the package. The analysis of
complete data, of data with missing values and of event data follow the same
pattern: a graph is selected, fitted and tested. The selected graph is in each case an
object of class \class{GraphInfer}, which also carries the record of the search and,
once fitted, the fit; the standard methods \fct{print}, \fct{summary}, \fct{plot},
\fct{logLik}, \fct{AIC}, \fct{BIC}, \fct{anova}, \fct{drop1} and \fct{add1} apply to
it. Table~\ref{tab:questions} lists the questions that the package answers and the
functions that answer them; the package help page \code{?GraphInfer} gives the same
index for all 48 exported functions.

\begin{table}[t]
\centering
\small
\begin{tabular}{L{62mm}L{58mm}l}
\toprule
Question & Functions & Section \\
\midrule
Which variables are conditionally independent given the others? &
\fct{minForest}, \fct{stepSearch}, \fct{searchGlobal} & \ref{sec:structure} \\
Which variables are close to a variable of interest? &
\fct{nearVertices}, \fct{subModel} & \ref{sec:local} \\
Is an edge, or a set of edges, needed? &
\fct{fitModel}, \fct{drop1}, \fct{add1}, \fct{anova}, \fct{lrtGauss} & \ref{sec:fitting}, \ref{sec:exact} \\
How strong are the partial associations of a Gaussian model? &
\fct{precisionMLE} & \ref{sec:precision} \\
Do covariates explain part of the dependence? &
\fct{residualise} & \ref{sec:covariates} \\
What if some values are missing? &
\fct{fitModelEM}, \fct{searchEM}, \fct{stepTreeEM}, \fct{imputeMissing}, \fct{searchMI} &
\ref{sec:missing} \\
Is the dependence that of irreversible accumulation on a tree? &
\fct{searchGlobal}, \fct{treeFromGraph}, \fct{accFit}, \fct{accTest} & \ref{sec:acctrees} \\
How would the procedures behave on data from a known model? &
\fct{rModel}, \fct{rModelData}, \fct{rAccTree}, \fct{rAccData} & \ref{sec:numerical} \\
\bottomrule
\end{tabular}
\caption{Questions and the functions that answer them. Table~\ref{tab:edit-query}
lists the functions for editing and querying a model.}
\label{tab:questions}
\end{table}

\section{Statistical framework}
\label{sec:framework}

\subsection{Decomposable conditional Gaussian models}
\label{sec:cg}

Let $V=\Delta\mathbin{\dot\cup}\Gamma$ index the variables, with $\Delta$ the discrete
and $\Gamma$ the continuous ones, and write $X=(I,Y)$ with $I=(I_\delta)_{\delta\in\Delta}$
and $Y=(Y_\gamma)_{\gamma\in\Gamma}$. Each $I_\delta$ takes values in a finite set
$\mathcal I_\delta$, so that $I$ takes values in $\mathcal I=\times_{\delta\in\Delta}
\mathcal I_\delta$, and $Y$ takes values in $\mathbb R^{|\Gamma|}$. The mixed
distributions are conditional Gaussian (CG) \citep{Lauritzen+Wermuth:1989,Lauritzen:1996}:
with $p(i)=\Pr(I=i)$, the conditional distribution $Y\mid I=i$ is
$N_{|\Gamma|}(\mu(i),\Sigma(i))$, so that for cells with $p(i)>0$ the joint density is
\[
 f(i,y)
 =
 p(i)(2\pi)^{-|\Gamma|/2}|\Sigma(i)|^{-1/2}
 \exp\left\{-\tfrac12[y-\mu(i)]^\top\Sigma(i)^{-1}[y-\mu(i)]\right\}.
\]
The model is \emph{homogeneous} when $\Sigma(i)=\Sigma$ for all cells with $p(i)>0$,
and \emph{heterogeneous} otherwise; in both, the conditional means may depend on the
cell. The purely continuous case is the Gaussian graphical model and the purely
discrete case the graphical log-linear model.

Let $G=(V,E)$ be an undirected graph. A missing edge imposes the restriction
$X_u\perp X_v\mid X_{V\setminus\{u,v\}}$; a present edge imposes none, and a
distribution in the model may satisfy further independences. The models are
restricted to \emph{decomposable} graphical models \citep{Edwards:2000,Lauritzen:1996},
whose graphs are chordal: every cycle of length four or more has a chord. The maximal
cliques $C_1,\ldots,C_m$ of a chordal graph can be arranged in a \emph{perfect
ordering}: with $H_j=C_1\cup\cdots\cup C_j$, each separator $S_j=C_j\cap H_{j-1}$,
$j=2,\ldots,m$, is contained in some $C_l$ with $l<j$. A path is \emph{forbidden} if its
end vertices are non-adjacent discrete variables and all its interior vertices are
continuous. A mixed graphical model is \emph{strongly decomposable} if its graph is
chordal and has no forbidden path \citep{Lauritzen:1996,Frydenberg+Lauritzen:1989};
equivalently, adding a new vertex adjacent to every discrete vertex gives a chordal
graph, the \emph{star-augmented} graph. \pkg{GraphInfer} works with decomposable
models for purely continuous or purely discrete data and with strongly decomposable
models when both types are present, homogeneous or heterogeneous.

\subsection{Likelihood factorisation and model dimension}
\label{sec:likelihood}

Let $x^{(1)},\ldots,x^{(n)}$ be independent observations of $X$. For $A\subseteq V$
write $\Delta_A=A\cap\Delta$, $\Gamma_A=A\cap\Gamma$, $g_A=|\Gamma_A|$ and
$\mathcal I_A=\times_{\delta\in\Delta_A}\mathcal I_\delta$. For $j\in\mathcal I_A$ let
$n_j$ be the number of observations whose discrete component in $\Delta_A$ equals
$j$, let $\mathcal I_A^+=\{j:n_j>0\}$ be the occupied cells, $K_A=|\mathcal I_A^+|$
(with $K_A=1$ and $n_1=n$ when $\Delta_A=\varnothing$), and let $\bar y_j$ be the mean
of $y_{\Gamma_A}$ in cell $j$. The maximised log-likelihood $\ell(A)$ of the saturated
marginal model on $A$ has the discrete part $\ell_D(A)=\sum_{j\in\mathcal I_A^+}n_j
\log(n_j/n)$. With $W_A=n^{-1}\sum_j\sum_{r\in j}(y^{(r)}_{\Gamma_A}-\bar y_j)
(y^{(r)}_{\Gamma_A}-\bar y_j)^\top$ the pooled within-cell covariance matrix, and
$W_{A,j}$ the covariance matrix within cell $j$, computed with divisor $n_j$,
\begin{align*}
\ell_{\mathrm{hom}}(A)&=\ell_D(A)-\frac n2\left\{g_A\log(2\pi)+\log|W_A|+g_A\right\},\\
\ell_{\mathrm{het}}(A)&=\ell_D(A)-\frac12\sum_{j\in\mathcal I_A^+}n_j
\left\{g_A\log(2\pi)+\log|W_{A,j}|+g_A\right\},
\end{align*}
when these matrices are non-singular. For a (strongly) decomposable graph with cliques
$C_1,\ldots,C_m$ in a perfect ordering and separators $S_2,\ldots,S_m$, the fitted
density is the product of the saturated marginal fits on the cliques divided by those
on the separators \citep{Lauritzen:1996,Frydenberg+Lauritzen:1989}, so that
\begin{equation}
\ell(G)=\sum_{j=1}^m\ell(C_j)-\sum_{j=2}^m\ell(S_j),
\qquad
d(G)=\sum_{j=1}^m d(C_j)-\sum_{j=2}^m d(S_j),
\label{eq:cliquesep}
\end{equation}
repeated separators counting with their multiplicities. This clique--separator
decomposition is the one likelihood calculation used throughout \pkg{GraphInfer}. The
dimension of a saturated marginal model counts occupied cells:
\begin{align*}
d_{\mathrm{hom}}(A)&=(K_A-1)+K_Ag_A+\frac{g_A(g_A+1)}2,\\
d_{\mathrm{het}}(A)&=(K_A-1)+K_A\left\{g_A+\frac{g_A(g_A+1)}2\right\},
\end{align*}
for the cell probabilities, the cell means and the covariance matrices.

The decomposition makes a single edge move local. If adding the edge $\{u,v\}$ to $G$
keeps the graph (strongly) decomposable, its separator $S$ is the set of common
neighbours of $u$ and $v$ (Section~\ref{sec:stepwise}), and with $Su=S\cup\{u\}$,
$Sv=S\cup\{v\}$ and $Suv=S\cup\{u,v\}$, the likelihood-ratio statistic and the change
of dimension are
\begin{equation}
D_{uv\mid S}=2\{\ell(Suv)+\ell(S)-\ell(Su)-\ell(Sv)\},
\qquad
q_{uv\mid S}=d(Suv)+d(S)-d(Su)-d(Sv).
\label{eq:fourterm}
\end{equation}
Both come from four saturated marginal calculations, without refitting the model. The
same identity serves the forest, stepwise and global searches.

In a heterogeneous model the covariance matrix of a cell can be estimated only if
$n_j>g_A$. By default \pkg{GraphInfer} regards the saturated marginal likelihood as
undefined when this fails in an occupied cell, and a search does not create such a
clique. The option \code{GraphInfer.sparseCells = "drop"} leaves such cells out of the
Gaussian part instead; this is not invariant to the scale of the continuous variables,
and is not part of the default procedure.

\paragraph{The occupied-cell dimension.}
Counting occupied cells rather than all possible ones avoids penalising parameters
that the data cannot estimate. The resulting $d(G)$ is the dimension associated with the fitted support: that of the
face of the parameter space that holds the maximum-likelihood estimate of $G$, since cells
whose clique margins are empty have estimated probability zero. It should be distinguished
from the nominal dimension of the model family. It equals the nominal dimension when
every cell is occupied, and with positive cell probabilities this is the case with
probability tending to one, so that the usual large-sample justifications of AIC and
BIC apply. With empty cells the estimate lies on the boundary, and the criteria are
then adaptive penalties based on the observed support rather than the standard AIC and
BIC; their regular justifications do not carry over, and singular models call for
other penalties \citep{Drton+Plummer:2017}. Moreover, the faces of nested models need
not be nested, so that $q_{uv\mid S}$ is not, in general, the number of degrees of
freedom of a test. For two binary variables observed only as $(0,0)$ and $(1,1)$, the
independence model has dimension 2 and the model with the edge dimension 1, so that
$q=-1$. For an edge between two discrete variables,
\begin{equation}
q_{uv\mid S}=m_S\sum_{j\in\mathcal I^+_{\Delta_S}}\bigl\{(a_j-1)(b_j-1)-z_j\bigr\},
\label{eq:qidentity}
\end{equation}
where $a_j$ and $b_j$ are the numbers of levels of $u$ and $v$ observed in cell $j$ of
$\Delta_S$, $z_j$ is the number of empty $(u,v)$ cells inside these observed margins,
and $m_S$ is the number of parameters per cell of the continuous part of $S$ ($1+g_S$
in the homogeneous and $1+g_S+g_S(g_S+1)/2$ in the heterogeneous model). The first term
is the usual degrees of freedom with empty rows and columns left out; the convention
subtracts in addition the internal zeros, as one would for structural zeros. In the
bipartite graph whose edges are the observed $(u,v)$ pairs of a cell, a term equals the
number of independent cycles minus the number of connected components plus one; it is
negative when the pairs fall into several blocks with few cycles, as in the example. For
an edge with a continuous endpoint $q_{uv\mid S}$ is never negative. In selection, an edge with $q_{uv\mid S}\le0$ never raises the criterion, whatever the
penalty, and lowers it when $D_{uv\mid S}>0$, which is always the case when
$q_{uv\mid S}<0$, since the observed pairs are then not those of independence; the extended BIC, whose edge term is never zero, limits this
(Section~\ref{sec:criteria}). Tests therefore use separate degrees of freedom,
described in Section~\ref{sec:fitting}.

\subsection{Model-selection criteria and conventions}
\label{sec:criteria}

The selection procedures minimise the penalised criterion
$\mathcal C_k(G)=-2\ell(G)+k\,d(G)$, $k\ge0$. The choices $k=0$, 2 and $\log n$ give
maximum likelihood, AIC and BIC \citep{Akaike:1974,Schwarz:1978}, specified by
\code{stat = "ML"} (\code{"LR"}, the name used in \pkg{gRapHD}, is an alias),
\code{"AIC"} and \code{"BIC"}, or by a numerical $k$. When the number of variables is
large relative to the sample size, BIC may select too many edges, and the package also
implements the extended BIC \citep{Chen+Chen:2008,Foygel+Drton:2010},
\[
\mathcal C^{\mathrm{EBIC}}_\gamma(G)=-2\ell(G)+(\log n)\,d(G)+4\gamma\,(\log|V|)\,|E|,
\]
with \code{stat = "EBIC"} ($\gamma=1/2$); \code{stat = c(k, kEdge)} gives
$-2\ell(G)+k\,d(G)+k_E|E|$ in general, and \fct{EBIC} evaluates the criterion of a fitted
model for any $\gamma$. The consistency results for the extended BIC were established
for Gaussian graphical models; for mixed, heterogeneous and sparse discrete models we
use it as a heuristic extension, whose behaviour Section~\ref{sec:study-missing}
reports. The case $k=0$ does not control complexity and is useful where the search
space is already restricted, as for maximum-likelihood trees.

By \eqref{eq:fourterm}, adding an admissible edge changes the criterion by
$-D_{uv\mid S}+kq_{uv\mid S}$ (plus $k_E$), and removing one by the opposite amount, so
that additions and removals are evaluated by the same local quantities. Numerical ties
need a convention: by default a change smaller in absolute value than
$\varepsilon_n=10^{-8}\max(1,n)$ is a tie. An edge is added only for a real gain beyond
this tolerance, whereas a removal is accepted at a tie, so that the sparser of two
indistinguishable models is preferred. The searches also accept a stricter threshold, a
minimum improvement before a move is made, without changing the criterion.
\fct{minForest} additionally accepts user-defined edge scores; these extend the search
mechanism and need not have a likelihood interpretation.

\section{Structure inference}
\label{sec:structure}

The structure-inference procedures search over classes of admissible graphs with the
criteria of Section~\ref{sec:criteria}: minimum forests, where the optimum is exact
(Section~\ref{sec:forest}); forward and backward stepwise search
(Section~\ref{sec:stepwise}); stochastic global search (Section~\ref{sec:global}); and
tools for local analysis around selected vertices (Section~\ref{sec:local}).

\subsection{Minimum forests}
\label{sec:forest}

Every forest is chordal, and in a forest the contribution of each edge does not depend
on the other edges. For a possible edge $\{u,v\}$, the statistics $D_{uv}$ and $q_{uv}$
of \eqref{eq:fourterm} with the empty separator are those of marginal independence of
$X_u$ and $X_v$. With the weight $w_{uv}^{(k)}=D_{uv}-kq_{uv}$, the criterion of a forest
$F=(V,E_F)$ is
\[
\mathcal C_k(F)=\mathcal C_k(G_0)-\sum_{\{u,v\}\in E_F}w_{uv}^{(k)},
\]
with $G_0$ the graph without edges, so that minimising $\mathcal C_k$ over forests is a
maximum-weight forest problem. This is the Chow--Liu construction
\citep{Chow+Liu:1968} and its extension to penalised criteria
\citep{Edwards+Abreu+Labouriau:2010}, solved by Kruskal's algorithm
\citep{Kruskal:1956}: edges are taken in order of decreasing weight and accepted when
they create no cycle. For $k=0$ the result is a maximum-likelihood spanning tree; with
AIC or BIC edges of non-positive weight are omitted, and the result is in general a
forest. For mixed data an edge is also rejected when it creates a forbidden path, and the
procedure yields the optimal strongly decomposable forest
\citep{Edwards+Abreu+Labouriau:2010}. The forest is thus one of the few structure
problems whose optimum is obtained without iterative search.

The function \fct{minForest} implements these procedures, with the criteria of
Section~\ref{sec:criteria} and case weights (the attribute \code{"weights"} of the data;
integer weights are equivalent to repeated rows). Edges may be excluded, and sets of
vertices forced to be cliques. With user-defined edge scores the result is a
maximum-weight spanning tree when the restrictions allow one: excluded edges can leave a
spanning forest, and forced cliques are included as given. The pairwise statistics are
computed for all pairs at once, for continuous, discrete and mixed data alike
(Appendix~\ref{app:structure}).

\subsection{Stepwise search}
\label{sec:stepwise}

For general decomposable models the criterion no longer splits into independent edge
contributions. The function \fct{stepSearch} moves through the admissible models by
adding or removing one edge at a time. In forward search, a non-edge is
\emph{add-eligible} if adding it keeps the graph chordal and, for mixed data, strongly
decomposable; the eligible edge with the largest decrease of $\mathcal C_k$ is added if
the decrease exceeds the threshold of Section~\ref{sec:criteria}, and the search stops
otherwise. Backward search removes \emph{remove-eligible} edges in the same way. Both
directions use the same likelihood, dimension and penalty, and, being greedy, both may
depend on the starting graph.

Decomposability makes the evaluation of a move local. If $u$ and $v$ are non-adjacent in
a chordal graph and $S$ is the set of their common neighbours, adding $\{u,v\}$ keeps
the graph chordal if and only if $S$ is complete and separates $u$ from $v$
\citep{Giudici+Green:1999}; $S$ is then the separator of \eqref{eq:fourterm}. An edge can
be removed while keeping the graph chordal if it lies in exactly one maximal clique $C$,
and then $S=C\setminus\{u,v\}$. For mixed data the same conditions are applied to the
star-augmented graph, which preserves strong decomposability.

Two forward engines are provided. The reference engine, \code{engine = "classic"},
scans all candidate non-edges at each step. The default, \code{engine = "fast"},
exploits the locality just described: an add-eligible edge joins two components or two
vertices at distance two, so that candidate scores can be cached by separator and
updated only where an addition changes the local structure
(Appendix~\ref{app:structure}). It reorganises the same greedy calculation, with the
same criterion, moves and tie-breaking, and therefore follows the same path
(Section~\ref{sec:benchmark-step}). With the default \code{join = FALSE} the search
keeps the components of the starting graph; \code{join = TRUE} also allows edges between
them. Edges may be excluded from addition, or protected from removal, by
\code{forbEdges}. For analyses written for \pkg{gRapHD}, \fct{stepw} is retained as a
forward-search interface with that package's calling conventions
\citep{Abreu+Edwards+Labouriau:2010}.

\subsection{Global search}
\label{sec:global}

A stepwise search follows one greedy path and may stop at a graph from which no single
addition or removal improves the criterion, although better graphs exist. The function
\fct{searchGlobal} explores the space of (strongly) decomposable graphs beyond such a
local optimum and keeps the best graph found. ``Global'' is meant in this algorithmic
sense: the result carries no certificate of optimality.

The default strategy is \emph{iterated local search} (ILS)
\citep{Lourenco+Martin+Stutzle:2003}. From the starting graph, a local optimum is first
reached by alternating forward and backward stepwise moves until neither improves the
criterion. A random perturbation, or \emph{kick}, then applies a sequence of admissible
single-edge additions and removals without requiring them to improve the criterion, and
the local optimisation is repeated. If the new local optimum is better than the best so
far it becomes the current graph; otherwise the current graph is kept. By default the
kick has a quarter of the number of edges of the first local optimum
(\code{kickFrac = 0.25}), with a small minimum, and the search runs $\max(50,p)$ rounds;
both can be increased for a wider search. These defaults come from the studies of
Section~\ref{sec:study-global}. The local optimisations use the fast engine, which makes
many rounds affordable.

Two further strategies are available. Simulated annealing
\citep{Kirkpatrick+Gelatt+Vecchi:1983} accepts a move that worsens the criterion by
$\Delta>0$ with probability $\exp(-\Delta/T)$, with a decreasing temperature $T$; a
scheduled search proposes additions with high probability at first, to grow away from a
sparse start, and then balances additions and removals under the same acceptance rule.
Both were less efficient than ILS in our studies. All strategies end with a backward pass
that removes edges whose removal does not worsen the criterion. Results are reproducible
for a fixed seed; agreement of runs from several seeds is useful evidence that the search
has explored the relevant part of the space, and disagreement a sign that more effort is
needed.

\subsection{Local analysis}
\label{sec:local}

When the question concerns a few variables of interest, a search over all variables may
be unnecessary. \fct{nearVertices} returns the vertices within a given graph distance of
a set of vertices, for instance in a minimum forest, and \fct{subModel} restricts a
model to a set $A$ of vertices, giving the induced subgraph $G[A]$. The two uses of a
subset differ in meaning. A model selected on the variables in $A$ alone describes the
marginal distribution of $X_A$: its missing edges are conditional independences given
the other variables in $A$, not given all of $V$, and marginalising over
$V\setminus A$ may add dependences. The induced subgraph of a model selected on all
variables keeps the conditioning on all of them. A forest neighbourhood is therefore a
screening device, not an estimated Markov blanket; examining successively larger
neighbourhoods and checking the stability of the local structure is a useful diagnostic,
but does not replace a full analysis when conditioning on all variables matters.

\section{Model fitting and inference}
\label{sec:inference}

\subsection{Fitting and testing a given model}
\label{sec:fitting}

The function \fct{fitModel} fits a decomposable model, selected by a search or given
directly (for example by \fct{makeModel}), by the decomposition \eqref{eq:cliquesep}. The
result remains an object of class \class{GraphInfer}; its component \code{fit} holds the
deviance $-2\ell(G)$, the dimension (including any offset from a covariate adjustment,
Section~\ref{sec:covariates}) and the number of observations. The methods \fct{logLik},
\fct{AIC} and \fct{BIC} return the corresponding quantities, \fct{summary} the cliques,
separators, search record and fit, and \fct{plot} the graph. \fct{fitModel} requires
complete observations; incomplete data are treated in Section~\ref{sec:missing}.

Nested models are compared by likelihood-ratio tests: \fct{anova} for given models, and
\fct{drop1} and \fct{add1} for every single-edge deletion or addition that keeps the
graph admissible, which are the moves of the stepwise searches. Each table reports two
kinds of degrees of freedom. \code{Df} is the difference of the dimensions
\eqref{eq:cliquesep}, which enters the criteria. \code{Test Df} is the number of degrees
of freedom of the test: the two models are compared on the same support, the cells with
positive estimated probability under the smaller model $G_0$, and \code{Test Df} is the
dimension of the larger model restricted to these cells minus that of $G_0$, in the
spirit of \citet{Fienberg+Rinaldo:2012}. It is never negative. For a single edge between
two discrete variables it is $m_S\sum_j(a_j-1)(b_j-1)$, the first term of
\eqref{eq:qidentity}; for an edge with a continuous endpoint it equals \code{Df}. The
chi-squared p-value uses \code{Test Df}, and is not given when it is zero, since the data
then carry no information on the edge. For several edges \fct{anova} computes the ranks
of the parameter designs on the support of $G_0$, which needs the data (argument
\code{data}). In the example of Section~\ref{sec:likelihood} the test has one degree of
freedom.

A model can also be edited and queried directly (Table~\ref{tab:edit-query}).
\fct{addEdge}, \fct{deleteEdge} and \fct{addClique} return a new model, fitted when the
data are supplied; several edits are applied at once, and only the final graph must be
(strongly) decomposable, since no edges are added automatically.

\begin{table}[t]
\centering
\small
\begin{tabular}{ll}
\toprule
Functions & Purpose \\
\midrule
\fct{addEdge}, \fct{deleteEdge}, \fct{addClique} & Edit the graph of a model \\
\fct{drop1}, \fct{add1} & Test single-edge moves \\
\fct{isDecomposable} & Decomposability and strong decomposability \\
\fct{separates} & Separation, and hence conditional independence \\
\fct{isCollapsible} & Collapsibility onto a set of vertices \\
\fct{adjMatrix}, \fct{edgeNames}, \fct{ripDecomp} & Adjacency, edges, cliques, separators \\
\fct{compareGraphs}, \fct{compareTrees} & Compare graphs, or accumulation trees \\
\fct{rModel}, \fct{rModelData} & Random models, and data from them \\
\bottomrule
\end{tabular}
\caption{Functions for editing, querying, comparing and simulating models.}
\label{tab:edit-query}
\end{table}

\subsection{Gaussian concentration-matrix estimation}
\label{sec:precision}

In a Gaussian graphical model with graph $G$, a missing edge $\{u,v\}$ is the zero
$K_{uv}=0$ of the concentration matrix $K=\Sigma^{-1}$. For a decomposable model the
maximum-likelihood estimate has the closed form \citep{Lauritzen:1996}
\[
\widehat K=\sum_{j=1}^m[S_{C_j}^{-1}]^0-\sum_{j=2}^m[S_{S_j}^{-1}]^0,
\]
where $S$ is the empirical covariance matrix with divisor $n$, $S_A$ its submatrix on
$A$, and $[\,\cdot\,]^0$ embeds a matrix in the rows and columns of $A$, with zeros
elsewhere; this is the clique--separator decomposition again. Only the clique and
separator matrices are inverted, so that $\widehat K$ exists when $n>|C_j|$ for every
clique and these matrices have full rank, even if $S$ itself is singular because $p>n$.
The function \fct{precisionMLE} computes $\widehat K$ for continuous decomposable models
(Section~\ref{sec:example-gauss} shows the partial correlations derived from it).

\subsection{Exact likelihood-ratio tests}
\label{sec:exact}

The chi-squared reference of \fct{anova} and \fct{drop1} is asymptotic. For nested
decomposable Gaussian models the likelihood-ratio statistic has an exact distribution
\citep[Propositions~5.12 and~5.14]{Lauritzen:1996}. Let $G_0\subseteq G_1$ be
decomposable on the same continuous variables, with estimated concentration matrices
$\widehat K_0$ and $\widehat K_1$, and $B=\det\widehat K_0/\det\widehat K_1$, so that the
deviance is $\Lambda=-n\log B$. The graph $G_0$ is reached from $G_1$ by a sequence
$G_1=G^{(0)}\supset G^{(1)}\supset\cdots\supset G^{(J)}=G_0$ of decomposable graphs, each
obtained from the previous one by removing one edge \citep[Lemma~2.21]{Lauritzen:1996};
the edge removed at step $i$ lies in a unique maximal clique $C_i$ of $G^{(i-1)}$, the
graph from which it is removed. Under $G_0$,
\[
B\overset{d}{=}\prod_{i=1}^JB_i,
\qquad
B_i\sim\operatorname{Beta}\left(\frac{n-|C_i|}2,\frac12\right)\ \text{independent}.
\]
The sequence is a device for factorising the statistic, not an assumption, and need not be
unique. The function \fct{lrtGauss} builds it, computes $B$, $\Lambda$ and both the exact
and the chi-squared ($\chi^2_J$) p-values. For one edge the exact p-value is the Beta
probability $\Pr\{\operatorname{Beta}(\alpha_1,1/2)\le B\}$; for several, the exact null
distribution is evaluated by Monte Carlo (\fct{QprodBeta}). With $k$ of the $N$
(\code{nsim}) simulated values at least as extreme as the observed one, the reported
p-value is $(1+k)/(N+1)$, which is itself a valid p-value for every $N$: under the null
hypothesis $\Pr\{(1+k)/(N+1)\le\alpha\}\le\alpha$, since the observed value is exchangeable
with the simulated ones \citep[Chapter~4]{Davison+Hinkley:1997}. Every Monte Carlo
p-value of \fct{lrtGauss} is computed in this way; it is at least $1/(N+1)$, and the output
says when no simulated value is as extreme.

The test extends to mixed models; edges between two discrete variables are treated at
the end of this section. Both models must be strongly decomposable, and the edges are removed one at a
time through strongly decomposable graphs; $\Lambda$ is the sum of the step deviances
and $B=\exp(-\Lambda/n)$. A step removes $\{u,v\}$ from its clique $C$, with
$S=C\setminus\{u,v\}$. Three facts, proved in Appendix~\ref{app:exactproofs}, make the
test exact. First, the step deviance is that of the regression of a continuous endpoint
$w$ on the cells of $\Delta_S$ and on $Y_{\Gamma_S}$, with and without the other
endpoint, which enters as a continuous regressor or through the cells of a discrete
variable (Proposition~\ref{prop:onestep}). Second, $w$ can be chosen so that its side
of the graph without the edge contains no discrete variable (Lemma~\ref{lem:side}); by
the Markov property, conditionally on the discrete data the step is then a normal linear
model whose null distribution is free of parameters. Third, the steps are independent,
by Basu's theorem (Proposition~\ref{prop:chain}).

In a homogeneous model the ratio of the residual sums of squares of step $i$ satisfies
\begin{align*}
B_i&\sim\operatorname{Beta}\left(\frac{n-K_{C}-g_{C}+1}2,\frac12\right)
&&\text{(continuous edge)},\\
B_i&\sim\operatorname{Beta}\left(\frac{n-K_{C}-g_{S}}2,\frac{K_{C}-K_{S}}2\right)
&&\text{(discrete--continuous edge)},
\end{align*}
with $K_A$ and $g_A$ as in Section~\ref{sec:likelihood}, and $B$ is the product of the
independent $B_i$. For one continuous edge this is the $t$ test of the other endpoint in
the regression, and for one discrete--continuous edge the corresponding $F$ test. These
single-edge tests are classical: the reduction to a regression is due to
\citet{Frydenberg+Lauritzen:1989}; \citet[Proposition~6.24 and pp.~192--194]{Lauritzen:1996}
reduces a single-edge test to the saturated model of its clique and gives the exact null
distributions, conditionally on the counts, in homogeneous and heterogeneous models; the
$F$ tests are available in MIM \citep[Section~5.3]{Edwards:2000}; and the Beta laws for
homogeneous models, with the parameters above, are derived by \citet{Tur+Castelo:2011} and
\citet{Tur+Roverato+Castelo:2014}, whose package \pkg{qpgraph} implements them. For
several edges, \citet{Frydenberg+Lauritzen:1989} factorise the likelihood ratio along a
removal sequence and note that the factors are in general independent only asymptotically;
\citet[Proposition~6.27]{Lauritzen:1996} states that, conditionally on the counts, the
factors of the edges with a continuous endpoint are independent, each with a distribution
of Box type, and outlines the proof. Proposition~\ref{prop:chain} gives a complete proof,
through strongly decomposable removal sequences, with the parameters stated in terms of
occupied cells; this gives the exact tests for several edges, in homogeneous and
heterogeneous models, that \fct{lrtGauss} implements. In a
heterogeneous model each cell $j$ of $\Delta_S$ has its own regression and variance, and
the deviance of a step is
\[
\sum_j\left\{n_j\log\frac{\sum_lR_{jl}+E_j}{n_j}-\sum_ln_{jl}\log\frac{R_{jl}}{n_{jl}}\right\},
\]
where $l$ runs over the occupied levels of a discrete endpoint within cell $j$, or over a
single group for a continuous edge. Under the null hypothesis $R_{jl}\sim\chi^2_{n_{jl}-c}$
and $E_j\sim\chi^2_{e_j}$ are independent, with $c=g_S+2$ and $e_j=1$ for a continuous
edge and $c=g_S+1$ and $e_j=(m_j-1)(g_S+1)$ for a discrete--continuous edge with $m_j$
occupied levels in cell $j$ (in units of the cell variance, which cancels). This
distribution is simulated, and every group needs more than $c$ observations.

An edge between two discrete variables can be removed from a strongly decomposable graph
only if its clique contains no continuous vertex, since a continuous vertex of the clique
would join the two endpoints by a forbidden path; its step deviance is the deviance of
$u\perp v\mid S$ in the contingency table of the clique. Its null distribution depends on
the discrete probabilities, and conditioning on the discrete data would leave nothing to
test. Under $G_0$ the discrete variables follow the decomposable log-linear model of $G_0$
restricted to them, whose parameters are variation independent of those of the conditional
Gaussian part, and the test conditions on its sufficient statistic, the margins of its
cliques, as the exact conditional tests for contingency tables do
\citep[Section~4.4.3]{Lauritzen:1996}. The edges with a continuous endpoint are removed
first and those between discrete variables last \citep[Lemma~2.22]{Lauritzen:1996}. The
null distribution of $\Lambda$ is simulated as outlined by \citet[p.~196]{Lauritzen:1996}:
discrete data from their conditional distribution given the margins, by Patefield's
algorithm \citep{Patefield:1981} applied clique by clique (within each cell of a separator,
the observed values of the new variables are assigned to the observations in random
order); the deviances of the discrete steps from these data; and, given them, the steps
with a continuous endpoint from the laws above, with the occupied cells and their counts of
the simulated data. Proposition~\ref{prop:discrete} shows that the resulting p-value is
exact conditionally on the margins. In purely discrete models this is the exact
conditional test of nested decomposable log-linear models, and for one edge without
conditioning variables the conditional test of Fisher with the deviance as its statistic.
Since the deviance takes few values, the p-value counts ties as at least as extreme.

All these null distributions are exact conditionally on the discrete data, or on their
margins under $G_0$ when edges between discrete variables are removed, so that the
ideal p-values, computed from them, are uniform under the null hypothesis, conditionally
and hence unconditionally (super-uniform for the discrete statistics, whose distributions
have atoms); when the tail probability is evaluated by Monte Carlo, the
reported p-value estimates the ideal one, with precision set by \code{nsim}. The
conditions for the test to apply depend on the discrete data only (simulated data sets
that fail them are left out, which conditions on an event of the discrete data). Like
every test of a fixed hypothesis, they do not account for the selection of a graph on the
same data (Section~\ref{sec:discussion}). Their level is examined in
Section~\ref{sec:study-exact}.

\subsection{Covariate adjustment}
\label{sec:covariates}

The models above describe dependence after allowing for the means. When the means
depend on explanatory variables such as treatment, block or site, responses affected by
the same covariates appear associated through their common mean structure. The function
\fct{residualise} (alias \fct{residualize}) removes this structure. With $Y$ the
$n\times p$ matrix of continuous responses and $X$ an $n\times q$ design matrix with an
intercept, common to all responses, the model is $Y=XB+E$ with independent rows
$E_i\sim N_p(0,\Sigma)$. The maximum-likelihood estimate of the means is then ordinary
least squares, $\widehat B=(X^\top X)^{-1}X^\top Y$, and inference on the covariance
structure can be based on the residuals $Y-X\widehat B$. \fct{residualise} builds $X$
from an \proglang{R} formula, so that factors, interactions and continuous covariates may
enter.

The graphical-model dimension already counts one mean per response; the design adds
$d_{\mathrm{off}}=p(q-1)$ parameters, stored as the attribute \code{dfOffset}. Given to a
search, it is carried by the resulting models, and the fitted dimension becomes
$d(G)+d_{\mathrm{off}}$. For a fixed design the offset is common to all graphs and does not
affect their comparison, but the reported dimensions and criteria count the mean
parameters. A missing edge between two responses then means conditional independence
given the other responses and the covariates; the covariates are not vertices. The
interpretation rests on the mean model and on a covariance common to all covariate
values. The exact test of Section~\ref{sec:exact} takes the design into account: the
residual cross-products have $n-q$ degrees of freedom instead of $n-1$, and the Beta
shapes become $(n-q+1-|C_i|)/2$ \citep{Anderson:2003}. \fct{lrtGauss} takes $q$ from the
argument \code{q} or from \code{dfOffset}; the exact test for mixed models does not yet
accept a covariate design.

\section{Inference with missing values}
\label{sec:missing}

With missing observations, \pkg{GraphInfer} distinguishes three tasks: fitting a given
graph by EM (\fct{fitModelEM}, Section~\ref{sec:em}), selecting a graph by structural EM
(\fct{searchEM}, Section~\ref{sec:sem}), and completing the data by imputation
(\fct{imputeMissing}, \fct{searchMI}, Section~\ref{sec:mi}). The missingness is assumed
to be ignorable, in particular missing at random with distinct parameters, so that
inference can be based on the observed-data likelihood without modelling the
missingness \citep{Little+Rubin:2019}.

\subsection{Maximum likelihood fitting by EM}
\label{sec:em}

For a fixed graph, the EM algorithm \citep{Dempster+Laird+Rubin:1977} alternates between
the conditional expectations of the complete-data sufficient statistics given the
observed data and the complete-data maximum-likelihood estimates computed from them; this
is the formulation of \citet{Lauritzen:1995} for graphical association models. The
likelihood maximised is that of the observed data. For a Gaussian model and an observation
with observed and missing coordinates $O$ and $M$, the missing coordinates are, at the
current $(\mu,\Sigma)$, Gaussian with
\begin{equation}
E(Y_M\mid Y_O)=\mu_M+\Sigma_{MO}\Sigma_{OO}^{-1}(Y_O-\mu_O),
\qquad
\operatorname{Var}(Y_M\mid Y_O)=\Sigma_{MM}-\Sigma_{MO}\Sigma_{OO}^{-1}\Sigma_{OM}.
\label{eq:condgauss}
\end{equation}
Both enter the expected statistics: the conditional covariance contributes to the
expected sums of squares and products, which a conditional mean alone would miss. The
M-step applies the decomposable estimator of Section~\ref{sec:precision} to the expected
mean and covariance. For discrete and mixed models, each discrete cell $c$ compatible with
the observed discrete values of a row receives a posterior weight proportional to
$p(c)\,\phi(y_O;\mu_O(c),\Sigma_{OO}(c))$, and within each cell the conditional moments of
the missing continuous coordinates follow \eqref{eq:condgauss}; the M-step applies the
closed-form estimates of the decomposable CG model \citep{Frydenberg+Lauritzen:1989} to the
expected cell counts and moments. \citet{Lauritzen:1995} computes the E-step by probability
propagation on a junction tree. \pkg{GraphInfer} holds mixed data, and discrete data with
at most $10^5$ cells, on the full table of discrete cells; mixed models are limited to
this size. Larger discrete tables are held as clique and separator tables, the E-step
listing the completions of each incomplete row and using propagation only for rows with
very many missing values. All engines give the same iterates (Appendix~\ref{app:structure}).

\fct{fitModelEM} stops when the relative change of the observed-data log-likelihood is
below a tolerance, and reports the trace, the convergence and the estimates; \fct{logLik},
\fct{AIC} and \fct{BIC} refer to the observed data. In discrete and mixed models the
dimension counts the cells with positive expected count. EM approaches a maximum with
zero cell probabilities only slowly, so that a cell can keep a tiny expected count and
still count as occupied, which changes the BIC by a multiple of $\log n$. After
convergence, cells with negligible expected count are therefore fixed at zero and EM is
run again, the result being kept if the log-likelihood does not drop; this is a numerical
device that makes the reported dimension agree with the limit of the iterations, not an
estimate of which cells are truly empty. Rows with nothing observed are discarded with a
warning, and without missing values \fct{fitModelEM} gives the fit of \fct{fitModel}.

\subsection{Structure inference by structural EM}
\label{sec:sem}

Structural EM does not search on a completed data set. In each round it searches on a
weighted pseudo-sample that reproduces exactly the conditional expectations, under the
current fit, of the complete-data sufficient statistics; the selected graph is then
refitted by EM and judged by the observed-data criterion. This is the principle of
\citet{Friedman:1997,Friedman:1998}, implemented by \fct{searchEM}.

Let $G_t$ be the graph at round $t$, with estimate $\widehat\theta_t$, and for a candidate
graph $G$ let $Q_t(G,\theta_G)=E_{\widehat\theta_t,G_t}\{\ell_c(G,\theta_G;X_{\mathrm{com}})
\mid X_{\mathrm{obs}}\}$. The expected criterion of round $t$ is
$\widetilde{\mathcal C}_{k,t}(G)=-2\max_{\theta_G}Q_t(G,\theta_G)+k\,d(G)$. The complete-data
log-likelihood of a CG model is linear in its sufficient statistics, the cell counts,
within-cell sums and cross-products, so that $\max_\theta Q_t(G,\theta)$ is the
complete-data maximised log-likelihood evaluated at their expectations. These
expectations are passed to the complete-data searches through a weighted
\emph{pseudo-sample}. For a discrete cell $c$ with expected count $N_c$, conditional mean
$m_c$ of its $q$ continuous variables and expected covariance
$S_c=V_c\operatorname{diag}(\lambda_{c1},\ldots,\lambda_{cq})V_c^\top$, the pseudo-sample
contains the $2q$ points
\[
m_c\pm\sqrt{q\lambda_{cj}}\,v_{cj},\qquad j=1,\ldots,q,
\]
each with weight $N_c/(2q)$, where $v_{cj}$ is the $j$th column of $V_c$; their total
weight, weighted mean and weighted covariance are exactly $N_c$, $m_c$ and $S_c$. This is
the symmetric point set of the unscented transform \citep{Julier+Uhlmann:2004}. Applied to
every cell of the full table (one weighted point per cell for discrete data), it reproduces
the expected statistics of every vertex set, since these are sums of the cell statistics,
so that every saturated marginal likelihood \eqref{eq:fourterm} computed on the
pseudo-sample is exact for the expected criterion. Without a full table, the weighted
completions themselves form the pseudo-sample, and rows treated by propagation contribute
expected tables computed on demand.

The searches of Sections~\ref{sec:structure} and~\ref{sec:acctree-search} are then applied
unchanged: \code{search = "stepwise"} (default) alternates forward and backward
\fct{stepSearch} until the graph is stable; \code{"forward"} and \code{"backward"} use one
direction (also \fct{stepSearchEM}); \code{"global"} uses \fct{searchGlobal}; \code{"forest"}
(also \fct{stepTreeEM}) selects a forest by Kruskal's algorithm, which is exact for the
expected criterion; and \code{compatible = TRUE} (also \fct{accSearchEM}) keeps every search
within the accumulation-tree compatible graphs. The selected graph $G_{t+1}$ is refitted by
\fct{fitModelEM}, and its observed-data criterion $\mathcal C_k^{\mathrm{obs}}(G)=
-2\ell_{\mathrm{obs}}(G)+k\,d(G)$ is evaluated, with $n$ the number of rows with at least one
observed value. The rounds end when the graph no longer changes, the observed-data
criterion stops improving, or a maximum number of rounds is reached. The final model is
fitted to the incomplete data, so that its likelihood, AIC and BIC refer to the observed
data.

Friedman's argument guarantees that an improvement of the expected criterion improves the
observed-data criterion when the dimension is fixed, and the EM refit only increases
$\ell_{\mathrm{obs}}$. Here the dimension counts cells with positive expected count, and
after the refit of Section~\ref{sec:em} these can change between rounds; the check that the
observed-data criterion improved is therefore part of the method, not only a numerical
safeguard.

The same property that makes the pseudo-sample exact explains a weakness. The expectations
are taken under $G_t$, so that a conditional independence of $G_t$ is built into the
filled-in statistics, and the association of an absent edge is attenuated in proportion
to the missing information: the rounds tend to confirm the current graph. By default
\fct{searchEM} therefore runs two sequences of rounds. The first, under the lighter
criterion \code{prestat} (AIC), reaches a denser graph, from which the rounds under
\code{stat} start; the observed-data criterion decreases along the second sequence. The
first sequence runs only when its penalties are lighter, and \code{prestat = NULL} gives a
single sequence. Section~\ref{sec:study-missing} reports the gain and its cost. As with the
complete-data searches, the result is a local optimum, which may depend on the starting
graph and, with global search, on the seed. The confirmation effect also keeps edges, so that
a search can end at a graph that another graph improves on the observed-data criterion
(Section~\ref{sec:missing-example}); results under several criteria are worth comparing on
that criterion.

\subsection{Imputation and multiple-imputation structure inference}
\label{sec:mi}

Completed data sets are useful for prediction, display and for assessing the stability
of a selected structure. \fct{imputeMissing} completes the data from a model fitted by
\fct{fitModelEM} or \fct{searchEM}: by default, a missing continuous value is replaced by
its conditional expectation given the observed values of its row, and a missing discrete
value by its most probable level, using the posterior probabilities of the cells in the
mixed case. Such a completion is a prediction: analysed as if it were observed, it
understates the variability. With \code{draw = TRUE} the missing values are drawn from
their conditional distribution \eqref{eq:condgauss}, after drawing a compatible cell in the
discrete and mixed cases, and repeated draws give multiple imputations \citep{Rubin:1987}.
With \code{proper = TRUE} (default) each imputation is drawn from a refit of the model to a
bootstrap resample of the rows, which adds the uncertainty of the estimates
\citep{Efron:1994,Honaker+King:2010}; this is not identical to Rubin's posterior
predictive construction.

\fct{searchMI} selects a graph on each of $m$ completed data sets (a forest, then
stepwise or global search) and records the selection frequency $\widehat\pi_e$ of each
edge, a measure of stability across imputations, not a posterior probability. The edges
with $\widehat\pi_e\ge\tau$ (\code{threshold}, default $0.5$) are added in decreasing order
of frequency as long as the graph stays (strongly) decomposable, revisiting the candidates
after each addition, since an edge can become admissible once another has supplied a
chord. The combined graph is fitted by \fct{fitModelEM}, so that its criteria are
comparable with those of \fct{searchEM}. The edge combination is specific to
\pkg{GraphInfer} and should be distinguished from Rubin's pooling rules. The conditional
independences of the imputation model are inherited, to some extent, by the completed
data: a sparse imputation model makes excluded associations hard to recover, and a denser
one is preferable. Table~\ref{tab:missing-procs} sets the three procedures side by side.

\begin{table}[t]
\centering
\small
\begin{tabular}{L{26mm}L{34mm}L{28mm}L{40mm}}
\toprule
Procedure & What the search sees & Under which model & Uncertainty about the missing values \\
\midrule
Single imputation & one completed data set & a fixed imputation model & lost (no conditional variance) \\
\fct{searchMI} & $m$ random completions & a fixed imputation model & represented by the variation between draws \\
\fct{searchEM} & the exact expected statistics & updated in each round & enters through the conditional second moments \\
\bottomrule
\end{tabular}
\caption{Three ways of selecting a graph with missing values. All three final graphs are
fitted to the observed data by \fct{fitModelEM}.}
\label{tab:missing-procs}
\end{table}

\section{Graphical models induced by accumulation trees}
\label{sec:acctrees}

A further class of models arises from the irreversible accumulation of events along a
branching history. The graph is then induced by a mechanism rather than chosen for
computational convenience. The theory is developed in \citet{Labouriau:AccumulationTrees};
here we state the results needed for the software, and concentrate on what the package
adds: reconstruction, restricted structure inference with tree moves, and the fitting and
testing of the full model.

\subsection{Graphs induced by accumulation trees}
\label{sec:acctree-map}
\label{sec:acctree-compat}

Let $T=(N,A)$ be a rooted tree with arcs directed away from the root, with one event per
arc; an event occurs at most once and is retained by every descendant. An observation
selects a node of the tree, possibly an intermediate one, and records the binary
indicators $Y_a$ of the events on the path from the root to it. Assume that every node of
the tree has strictly positive probability of being selected as the stopping node. Under
this non-degeneracy condition, and in the absence of observation errors, the
conditional-independence graph of the event indicators is the line graph of the underlying
undirected tree, $G(T)=L(T^u)$ \citep{Labouriau:AccumulationTrees}: two events are adjacent
exactly when their arcs share a node. The graph is chordal, its maximal cliques are the
events incident to the same node, and two cliques meet in at most one event. Conversely, a
connected graph $G$ with maximal cliques $C_1,\ldots,C_m$ is the graph of an accumulation
tree, with one event per arc, if and only if
\begin{itemize}
\item[(i)] $G$ is chordal,
\item[(ii)] $|C_a\cap C_b|\le1$ for $a\ne b$, and
\item[(iii)] every vertex lies in at most two maximal cliques;
\end{itemize}
equivalently, the compatible connected graphs are the line graphs of trees. The function
\fct{isCompatible} checks these conditions componentwise, so that a disconnected graph is
compatible with a forest of accumulation trees, as intermediate graphs of a search often
are.

For a compatible graph, \fct{treeFromGraph} reconstructs the tree: one node per maximal
clique, an edge carrying the event $s$ between the nodes of two cliques that meet in $s$,
and a pendant edge for each event in only one clique. Since $G(T)$ depends only on $T^u$,
every choice of root gives the same graph: the graph identifies the undirected tree and the
placement of the events, but not the root, which must be supplied to orient the tree. The
functions \fct{accTree} and \fct{rAccTree} construct given or random trees, \fct{rAccData}
simulates event data, \fct{inducedGraph} returns the graph of a tree, and \fct{plot} draws
it. The graph does not express the restrictions that irreversibility places on the support
of the joint distribution; Section~\ref{sec:acctree-full} takes them up.

\subsection{Restricted structure inference and tree moves}
\label{sec:acctree-search}

When accumulation on a tree is adopted as a modelling assumption, structure inference can
be restricted to compatible graphs. The restriction is part of the statistical model: a
markedly better graph outside the compatible class is evidence against the accumulation
hypothesis, which is how the graph-level test of Section~\ref{sec:acctree-full} uses it.
The function \fct{accSearch} performs a greedy search with the criterion
$\mathcal C_k$ in which a move is accepted only if the result is compatible. Such
single-edge searches meet an obstruction that is not merely the usual risk of a local
optimum. A node with four incident arcs gives the clique $K_4$; to build it one edge at a
time, the edge completing $K_4$ must be added to $K_4$ minus an edge, which consists of two
triangles sharing two vertices and violates condition (ii), and removing an edge of $K_4$
leads to the same graph. In general:

\begin{proposition}\label{prop:obstruction}
Let $G$ be compatible and let $C$ be a maximal clique of $G$ with $|C|\ge4$. Let $G'$ be
obtained from $G$ by adding one edge, removing one edge, or removing one edge and adding
another. If $G'$ is compatible, then $G'$ has a clique with at least four vertices.
\end{proposition}

\begin{proof}
If no edge of $C$ is removed, $C$ is complete in $G'$, and the maximal clique of $G'$
containing it has at least four vertices. Otherwise an edge $\{a,b\}\subseteq C$ is removed,
and the edge added, if any, is not $\{a,b\}$. In $G'$ the sets $C\setminus\{a\}$ and
$C\setminus\{b\}$ are complete, and no clique contains both, since $a$ and $b$ are not
adjacent. They therefore lie in distinct maximal cliques of $G'$, which share the at least
two vertices of $C\setminus\{a,b\}$. This violates condition (ii), so $G'$ is not
compatible.
\end{proof}

\begin{corollary}\label{cor:unreachable}
A compatible graph with a clique of four or more vertices cannot be reached from the empty
graph, or from any compatible graph whose cliques have at most three vertices, by
single-edge additions, removals or swaps that keep every intermediate graph compatible. In
terms of the tree, trees with a node incident to four or more arcs are not reachable.
\end{corollary}

For six and seven events complete enumeration shows the converse: every compatible graph
whose cliques have at most three vertices is reachable from the empty graph by single-edge
moves (4\,456 and 56\,932 graphs; Section~\ref{sec:study-compatible}).

The obstruction disappears when the search moves in the underlying forest instead of the
line graph. Every compatible graph is the line graph of a forest $F$ whose edges are the
events. For a node $x$ of $F$, the events incident to $x$ form a clique $C_x$ when
$\deg(x)\ge2$, an event joining two nodes of degree at least two is a singleton separator,
and an isolated event is a one-vertex component. With $c_k(A)=-2\ell(A)+k\,d(A)$, the
criterion of the line graph $G_F$ is
\[
\mathcal C_k(G_F)=\sum_{x:\deg(x)\ge2}c_k(C_x)
-\sum_{\substack{e=\{x,y\}\in F\\ \deg(x),\deg(y)\ge2}}c_k(\{e\})
+\sum_{\substack{e=\{x,y\}\in F\\ \deg(x)=\deg(y)=1}}c_k(\{e\}),
\]
the clique--separator decomposition in the coordinates of the forest, so that a local
change of the forest changes few terms. A \emph{tree move} detaches one endpoint of an
event-edge and reattaches it to another node or to a new node; it can regraft a subtree,
merge components or split a node, and may add and remove several graph edges at once.

\begin{proposition}\label{prop:connected}
Every compatible graph can be reached from every other by tree moves, through compatible
graphs.
\end{proposition}

\begin{proof}
Every forest is an admissible state, since its line graph is compatible. Repeatedly
detaching an endpoint of an event-edge and placing it at a new node transforms any forest
into the forest of isolated event-edges; tree moves are reversible, so this forest can in
turn be transformed into any other.
\end{proof}

The compatible mode of \fct{searchGlobal} (\code{compatible = TRUE}) uses tree moves by
default, within the iterated local search of Section~\ref{sec:global}: improving tree moves
until a local optimum, then a perturbation by random tree moves, and again a descent.
Connectivity makes every compatible graph reachable, but the descent accepts only improving
moves and the kicks give no convergence guarantee, so that success is an empirical matter
(Section~\ref{sec:study-compatible}). Tree moves require all variables to be discrete or all
continuous; for mixed data the compatible search uses single-edge moves supplemented by
swaps, which remove one edge and add another. \fct{accSearch} remains useful for local
exploration and for simplifying a compatible model.

\subsection{The full accumulation model and tests of the accumulation hypothesis}
\label{sec:acctree-full}

The compatible graphical model is a relaxation of the accumulation mechanism. It does not
express the support restrictions of irreversibility, under which an event implies the events
above it and events on different branches exclude each other, and it assumes that events are
recorded without error. The function \fct{accFit} fits the full model. A lineage stops at
node $z$ of the rooted tree with probability $\pi_z$, $\sum_z\pi_z=1$; the events on the path
from the root to $z$ are present and the others absent, which defines latent indicators
$x_a(z)$; each recorded event is then flipped independently with probability
$\varepsilon_a$:
\[
P(Y=y)=\sum_{z\in N}\pi_z\prod_{a\in A}\varepsilon_a^{|y_a-x_a(z)|}(1-\varepsilon_a)^{1-|y_a-x_a(z)|}.
\]
The argument \code{noise} chooses separate noise probabilities (default), a common one, or
none; missing values are assumed to be missing at random and drop out of the product.

Accumulation of irreversible events under observation error is the subject of a large
literature on tumour progression. In oncogenetic trees \citep{Desper+others:1999} each arc
is retained independently with its own probability and an observation records the events
reachable from the root, so that events on different branches can occur together;
\citet{Szabo+Boucher:2002} added false-positive and false-negative rates common to all
events. Conjunctive Bayesian networks replace the tree by a partial order
\citep{Beerenwinkel+Eriksson+Sturmfels:2007}, and the hidden conjunctive Bayesian network
of \citet{Gerstung+others:2009} adds an observation error common to all events, with
exponential waiting times, fitted by EM. The model here differs in its sampling scheme: a
unit follows a single lineage, which stops at a node, so that events on different branches
exclude each other and the noise-free support is the set of root paths rather than of
rooted subtrees. It is a latent-class model with one class per node and a deterministic
profile per class, perturbed by symmetric flips with event-specific or common
probabilities. What is specific to \pkg{GraphInfer} is this stopping-node likelihood, its
fitting together with the search over trees, and its comparison with decomposable
graphical models, which rests on the line-graph form of the induced graph.

The parameters are estimated by EM with the stopping node as a latent class; probabilities
estimated at zero are fixed at zero as in Section~\ref{sec:em}, and the dimension counts the
positive node probabilities, minus one, and the positive noise probabilities. Unlike the graph, the full model can identify the root under suitable conditions, since
the support restrictions depend on the orientation of the tree; it does not always do so,
for instance when all lineages stop at the root and the events arise from noise alone. The
fit reports the profile log-likelihood over the roots. With separate noise probabilities the model contains mutual independence
(all lineages stopping at the root, $\varepsilon_a=P(Y_a=1)$), so that a large estimated
noise probability marks an event that the tree does not explain. \fct{eventCompatibility}
checks that pairs of events are nested or exclusive, as they are in a tree without noise,
and \fct{perfectPhylogeny} builds a rooted tree from the events, a perfect phylogeny in
the sense of \citet{Gusfield:1991} when all pairs are compatible. By default \fct{accFit}
starts from this tree and from the star, tries every root, and improves the tree by moving
an event with the events below it; the search is local, and \code{forest = TRUE} allows
several independent trees.

\begin{table}[t]
\centering
\small
\begin{tabular}{L{16mm}L{40mm}L{34mm}L{40mm}}
\toprule
Level & Null hypothesis & Bootstrap from & Main limitation \\
\midrule
\code{"graph"} (default) & noise-free dependence structure compatible with an accumulation
tree & the fitted compatible graphical model $\widehat G_{\mathrm{comp}}$, both searches
repeated & rejects under measurement noise; does not test the support restrictions \\
\code{"full"} & the accumulation mechanism, with its support restrictions and noise model &
the fitted accumulation model & liberal when many estimates lie on the boundary \\
\bottomrule
\end{tabular}
\caption{The two levels of \fct{accTest}.}
\label{tab:acctest-levels}
\end{table}

The function \fct{accTest} tests the accumulation hypothesis at the two levels of
Table~\ref{tab:acctest-levels}. The graph level starts from the graph
$\widehat G_{\mathrm{free}}$ selected without restriction and its best compatible subgraph
$\widehat G_{\mathrm{comp}}$, found by tree moves; both are the best graphs the searches
found, not certified optima. The reduction to a subgraph is needed because the complete
graph is compatible, being the line graph of a star, so that a comparison with the best
compatible graph overall would have no power. The full level compares the fitted
accumulation model with $\widehat G_{\mathrm{free}}$, or with the saturated model
(\code{unrestricted = "saturated"}), on the observed-data likelihood; with
\code{forest = TRUE} the null hypothesis is a forest. At both levels the statistics are the
difference of the criteria and a likelihood contrast, $2\{\ell(\widehat G_{\mathrm{free}})-
\ell(\text{null model})\}$. They are not referred to a chi-squared distribution: at the
graph level both graphs are selected on the data, at the full level the models are not
nested and the contrast can be negative, and even against the saturated model the
accumulation model is a latent-class model with boundary parameters. Their null
distribution is obtained by a parametric bootstrap: $B$ data sets are simulated from the
fitted null model, each row with the missingness pattern of a row of the data drawn at
random, both models are refitted to each, and for a statistic $T$ with bootstrap values
$T^*_b$ the p-value is $(1+\#\{T^*_b\ge T\})/(B+1)$.

This bootstrap is liberal when many estimates lie on the boundary, because the simulated
data are then sparser than data from the true model. At the full level the argument
\mbox{\code{fdb = TRUE}} adds the fast double bootstrap of \citet{Davidson+MacKinnon:2007}, at
twice the cost: each replicate also fits the accumulation model to its bootstrap data set,
simulates one further data set from that fit and computes $T^{**}_b$; with $p_1$ the bootstrap p-value and $Q$ the $1-p_1$ quantile of
the $T^{**}_b$, the p-value is $(1+\#\{T^*_b\ge Q\})/(B+1)$. Section~\ref{sec:study-acctest}
reports the calibration studies: they showed no substantial calibration problem in the
configurations with a well-determined fitted model, and a liberal test, improved but not
corrected in general by the fast double bootstrap, when many estimates lie on the boundary.
The procedure should not be regarded as calibrated beyond the configurations studied. The replicates are
independent and can be run in parallel through the argument \code{lapplyFun}.

\section{Using \pkg{GraphInfer}}
\label{sec:using}

This section shows the package at work on five data sets: continuous data with
covariates, mixed data, mixed data with missing values, gene-expression data with 1522
variables, and simulated event data. The \code{state.x77} data come with \proglang{R}; the
\code{wine}, \code{heartdisease} and \code{rnaseq} data come with \pkg{GraphInfer}, prepared from
the UCI Machine Learning Repository. Every result can be reproduced with the scripts of the
replication material, and the seeds are shown in the code.

\subsection{Continuous data with covariates}
\label{sec:example-gauss}

The data \code{state.x77} of the \proglang{R} package \pkg{datasets} give eight
characteristics of the 50 states of the USA in the 1970s
\citep{USCensus:1977}: population, income per head, illiteracy, life expectancy, murder
rate, share of high-school graduates, days with frost and area. Most of them differ
between the four census regions, so that two variables can appear associated only
through their regional means. The means are therefore adjusted for region before a
graph is selected:
\begin{CodeChunk}
\begin{CodeInput}
R> X <- as.data.frame(state.x77)
R> names(X) <- c("Pop", "Income", "Illit", "LifeExp", "Murder", "HSGrad",
+    "Frost", "Area")
R> R <- residualise(X, ~ region, data.frame(region = state.region))
R> m0 <- minForest(R)
R> m1 <- stepSearch(m0, R, dfOffset = attr(R, "dfOffset"))
R> m2 <- lapply(1:5, function(s) searchGlobal(m1, R, seed = s))
R> bic <- sapply(m2, function(m) BIC(fitModel(m, R)))
R> fit <- fitModel(m2[[which.min(bic)]], R)
\end{CodeInput}
\end{CodeChunk}
The region design adds $8\times3=24$ mean parameters, which the searches carry as
\code{dfOffset}. With them, the BIC forest has 7 edges and BIC 4331.7, the stepwise
search 8 edges and BIC 4327.0, and the global search, from five seeds, 11 edges and BIC
4317.7; four of the five seeds reached this value and the fifth 4319.3. The partial
correlations of the selected model come from the decomposable estimate of the
concentration matrix:
\begin{CodeChunk}
\begin{CodeInput}
R> P <- -cov2cor(precisionMLE(fit, R)); diag(P) <- 1
R> dimnames(P) <- list(names(R), names(R)); round(P, 2)
\end{CodeInput}
\begin{CodeOutput}
          Pop Income Illit LifeExp Murder HSGrad Frost Area
Pop      1.00   0.00 -0.36    0.00   0.46   0.00 -0.51 0.00
Income   0.00   1.00  0.00    0.00   0.00   0.41  0.00 0.31
Illit   -0.36   0.00  1.00    0.00   0.41  -0.39 -0.46 0.00
LifeExp  0.00   0.00  0.00    1.00  -0.62   0.00 -0.35 0.00
Murder   0.46   0.00  0.41   -0.62   1.00   0.00  0.02 0.00
HSGrad   0.00   0.41 -0.39    0.00   0.00   1.00  0.00 0.00
Frost   -0.51   0.00 -0.46   -0.35   0.02   0.00  1.00 0.00
Area     0.00   0.31  0.00    0.00   0.00   0.00  0.00 1.00
\end{CodeOutput}
\end{CodeChunk}
The zeros are the missing edges. Given the other variables and the region, the murder
rate is associated with life expectancy, population and illiteracy, and income only with
the share of graduates and the area. The edge Murder--Frost has a very small estimated partial correlation, 0.02, and cannot be
deleted on its own without destroying decomposability: it lies in the two cliques
$\{\mathrm{Pop},\mathrm{Illit},\mathrm{Murder},\mathrm{Frost}\}$ and
$\{\mathrm{LifeExp},\mathrm{Murder},\mathrm{Frost}\}$, and deleting it would leave the
chordless cycle LifeExp--Murder--Pop--Frost. It is the one edge that \fct{drop1} does not
list:
\begin{CodeChunk}
\begin{CodeInput}
R> drop1(fit, data = R, k = log(50))
\end{CodeInput}
\begin{CodeOutput}
Single-edge deletions (BIC with k = 3.912)
1 edge(s) not listed: deleting them alone leaves a graph that is not decomposable

                 Df -2logL    BIC    LRT Test Df  Pr(>Chi)
<none>              4118.2 4317.7
Pop - Illit       1 4126.7 4322.3  8.489       1 0.0035739 **
LifeExp - Frost   1 4129.1 4324.7 10.888       1 0.0009680 ***
Income - Area     1 4124.6 4320.2  6.409       1 0.0113529 *
\end{CodeOutput}
\end{CodeChunk}
(three of the ten rows shown). For the weakest edge the exact test of
Section~\ref{sec:exact} takes the four design columns into account through
\code{dfOffset}:
\begin{CodeChunk}
\begin{CodeInput}
R> lrtGauss(deleteEdge(fit, c("Income", "Area")), fit, R)
\end{CodeInput}
\begin{CodeOutput}
Exact LRT for nested Gaussian graphical models
  B = 0.879691   Lambda = 6.4092   df = 1
  p-value (exact, Beta) = 0.01691
  p-value (chi-square)    = 0.01135
\end{CodeOutput}
\end{CodeChunk}
With 50 observations the chi-squared p-value is about two thirds of the exact one, even
for a single edge.

\subsection{Mixed data}
\label{sec:workflow-example}

The data set \code{wine} of \pkg{GraphInfer} \citep{Aeberhard+Forina:1992}, from the UCI
Machine Learning Repository under its CC BY 4.0 licence, gives 13 chemical
measurements on 178 wines from three cultivars: alcohol (\code{Alch}), malic acid
(\code{Mlca}), ash, alcalinity of ash (\code{Aloa}), magnesium (\code{Mgns}), total phenols
(\code{Ttlp}), flavanoids (\code{Flvn}), nonflavanoid phenols (\code{Nnfp}),
proanthocyanins (\code{Prnt}), colour intensity (\code{Clri}), hue, OD280/OD315 of diluted
wines (\code{Oodw}) and proline (\code{Prln}). The variables are thus one factor, the
cultivar \code{Cult}, and 13 continuous variables. A first analysis takes a BIC forest as
an overview, extends it by forward stepwise search, and continues with a global search:
\begin{CodeChunk}
\begin{CodeInput}
R> library("GraphInfer")
R> data("wine", package = "GraphInfer")
R> m0 <- minForest(wine)
R> m1 <- stepSearch(m0, wine)
R> m2 <- searchGlobal(m1, wine, seed = 1)
\end{CodeInput}
\end{CodeChunk}
These calls use the default homogeneous conditional Gaussian model, in which the
covariance matrix is common to the cultivars. That assumption should be examined, and
the same searches are therefore repeated with \code{homog = FALSE}; the later searches
inherit the setting from their starting model:
\begin{CodeChunk}
\begin{CodeInput}
R> mh0 <- minForest(wine, homog = FALSE)
R> mh1 <- stepSearch(mh0, wine)
R> mh2 <- searchGlobal(mh1, wine, seed = 1)
R> fit <- fitModel(mh2, wine)
\end{CodeInput}
\end{CodeChunk}

The heterogeneous model is much better supported. The forest, stepwise and global
searches gave 13, 28 and 32 edges with BIC 6840.6, 6636.2 and 6596.4, against 13, 27 and
28 edges with BIC 6949.1, 6779.5 and 6779.0 under the homogeneous model: the best
heterogeneous graph is 182.6 units lower, in spite of more than twice the number of
parameters (137 against 65). The analysis therefore continues with this model. The
results also show why several seeds are useful: in the homogeneous case the global search
gains only 0.5 BIC units over the stepwise search, and only one of five seeds found that
graph, whereas in the heterogeneous case the gain is 39.8 units and all five seeds reach
the same value.

The route taken by a search matters. Table~\ref{tab:wine-routes} compares greedy
forward search, greedy backward search from the saturated model and the global
search, under BIC, AIC and the extended BIC. Forward search gave the same graph from
the empty graph as from the forest in all six cases. Under the heterogeneous model it
stops 11 to 40 units above the best value found, and backward search 57 to 181 units
above it. Backward search is hampered by strong decomposability: at its stopping
graph under BIC, only 19 of its 50 edges can be deleted singly. The global search was
started from each of the three greedy graphs. Under the heterogeneous model with BIC
or the extended BIC all three starts reached the same value, whereas in the other four
cases they differed by up to 9.6 units, so that several starts remain advisable.

\begin{table}[t]
\centering
\small
\begin{tabular}{lrrrrrrrr}
\toprule
 & \multicolumn{4}{c}{Homogeneous} & \multicolumn{4}{c}{Heterogeneous} \\
\cmidrule(lr){2-5}\cmidrule(lr){6-9}
Criterion & Best & Edges & Forward & Backward & Best & Edges & Forward & Backward \\
\midrule
BIC  & 6779.0 & 28 & 0.5  & 13.0 & 6596.4 & 32 & 39.8 & 181.0 \\
AIC  & 6550.4 & 43 & 19.8 & 5.1  & 6088.5 & 55 & 23.0 & 57.4 \\
EBIC & 6902.4 & 21 & 4.5  & 9.6  & 6748.6 & 22 & 11.2 & 99.5 \\
\bottomrule
\end{tabular}
\caption{Searches on the \code{wine} data. ``Best'' is the lowest value of the
criterion found, always by the global search, with the number of edges of that graph.
``Forward'' and ``Backward'' give the excess of the greedy searches over that value.}
\label{tab:wine-routes}
\end{table}

For comparison we ran the stepwise searches of \pkg{gRim} (version 0.3.4), among
decomposable models with penalties $\log n$ and 2. Its mixed model \code{mmod(\~{}.\^{}.)}
is homogeneous, with saturated dimension 132, so that the comparison belongs to the
homogeneous columns; its log-likelihoods and dimensions agree with those of
\pkg{GraphInfer}. Its graphs have 19 and 20 edges under BIC, and 25 and 44 under AIC,
with values 7.8 to 39.5 units above the best homogeneous values of
Table~\ref{tab:wine-routes}.

\begin{CodeChunk}
\begin{CodeInput}
R> summary(fit)
\end{CodeInput}
\begin{CodeOutput}
GraphInfer model
  vertices : 14 (13 continuous, 1 discrete)
  edges    : 32
  type     : mixed
  criterion: BIC
  search   : searchGlobal (ils)
  criterion change from start: -39.8358
  degree   : min 2, median 4, max 13
  cliques  : 11 (sizes 3-4), 10 separators
  -2 logL  : 5886.500  (df 137 = graph 137 + offset 0, n 178)
\end{CodeOutput}
\end{CodeChunk}

\begin{figure}[t]
\centering
\includegraphics[width=0.56\textwidth]{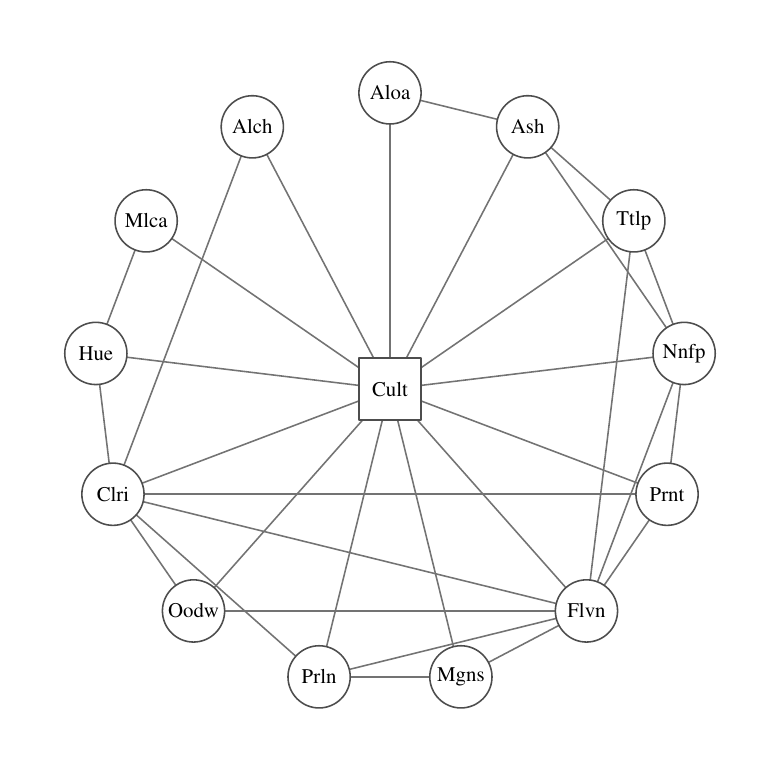}
\caption{Graph selected for the \code{wine} data by BIC under the heterogeneous
model. The square is the discrete variable. The layout is set in the replication
script; \code{plot(fit)} draws the same graph with an automatic layout.}
\label{fig:wine}
\end{figure}

The selected graph is shown in Figure~\ref{fig:wine}. The cultivar is adjacent to all
13 measurements, while the measurements are conditionally independent of one another
for most pairs: of their 78 pairs, 19 are edges. An edge between the cultivar and a measurement
is present when the mean, the variance or the regression on the separator differ
between cultivars.

The method \fct{drop1} tests the single-edge deletions that keep the graph strongly
decomposable. Of the 32 edges, 19 are eligible, and we show the four whose deletion
raises the BIC least:
\begin{CodeChunk}
\begin{CodeInput}
R> drop1(fit, data = wine, k = log(nrow(wine)))
\end{CodeInput}
\begin{CodeOutput}
            Df -2logL    BIC    LRT Test Df  Pr(>Chi)
<none>         5886.5 6596.4
Ash - Ttlp   3 5906.3 6600.7 19.811       3 0.0001858 ***
Clri - Oodw  3 5904.4 6598.8 17.938       3 0.0004531 ***
Clri - Hue   3 5905.3 6599.7 18.843       3 0.0002947 ***
Cult - Mlca  6 5919.0 6597.8 32.536       6 1.288e-05 ***
\end{CodeOutput}
\end{CodeChunk}
No eligible deletion lowers the BIC. The p-values here are chi-squared
approximations, whereas \fct{lrtGauss} gives the exact test of
Section~\ref{sec:exact}. In a heterogeneous model the exact
null distribution is simulated:
\begin{CodeChunk}
\begin{CodeInput}
R> lrtGauss(deleteEdge(fit, c("Clri", "Oodw")), fit, wine, seed = 1)
\end{CodeInput}
\begin{CodeOutput}
Exact LRT for nested mixed (heterogeneous) graphical models
  B = 0.904138   Lambda = 17.9377   df = 3
  p-value (exact, weighted sum of log chi-squares, MC) = 0.0009999
  p-value (chi-square)    = 0.0004531
\end{CodeOutput}
\end{CodeChunk}
The two p-values lead to the same conclusion, but the chi-squared one is about half
the exact one, even with 178 observations; the Monte Carlo standard error of the exact
value is about 0.0003. In a homogeneous model a one-edge test needs no simulation,
because its null distribution is a Beta distribution.

\subsection{Missing values}
\label{sec:missing-example}

The heart disease data \citep{Janosi+others:1989,Detrano+others:1989}, also from the UCI
Machine Learning Repository, describe 920 patients examined for coronary artery disease at
the Cleveland Clinic Foundation, the Hungarian Institute of Cardiology in Budapest, the
University Hospitals of Zurich and Basel, and the V.A. Medical Center in Long Beach. We use
the 14 attributes of published analyses: age, resting blood pressure (\code{BP}), serum
cholesterol (\code{Chol}), maximum heart rate (\code{MaxHR}) and ST depression induced by
exercise (\code{STdep}), which are continuous; sex, chest pain type (\code{Pain}), fasting
blood sugar above 120 mg/dl (\code{Sugar}), resting electrocardiogram (\code{ECG}),
exercise-induced angina, slope of the peak exercise ST segment, number of major vessels
coloured by fluoroscopy (\code{Vessels}) and thallium scan (\code{Thal}), which are
discrete; and the diagnosis, coded as presence or absence of disease. The centre is added as
a fifteenth variable. The data set \code{heartdisease} of \pkg{GraphInfer} contains these data,
read from the four processed files, with a recorded cholesterol or blood pressure of 0 taken
as missing (all Swiss cholesterol values are 0):
\begin{CodeChunk}
\begin{CodeInput}
R> data("heartdisease", package = "GraphInfer"); heart <- heartdisease
R> x <- colSums(is.na(heart)); x[x > 0]
\end{CodeInput}
\begin{CodeOutput}
     BP    Chol   MaxHR   STdep   Sugar     ECG  Angina   Slope Vessels
     60     202      55      62      90       2      55     309     611
   Thal
    486
\end{CodeOutput}
\end{CodeChunk}
Only 299 patients are complete, 297 of them from Cleveland, so that an analysis of the
complete rows is an analysis of one centre; and the rows differ: the disease is present in
60\% of the incomplete rows against 47\% of the complete ones. Most values are missing
because some centres did not record some examinations; with the centre in the model, the
analysis assumes that the values are missing at random given the observed variables. The
patients were referred for examination, so that the results illustrate the methods and are
not clinical findings.

The procedures of Section~\ref{sec:missing} are compared on the same observed-data criteria:
structural EM with BIC and with the extended BIC, a graph selected on the complete rows and
refitted to all rows by EM, and multiple imputation with the structural EM graph as
imputation model:
\begin{CodeChunk}
\begin{CodeInput}
R> sem <- searchEM(NULL, heart, search = "global", seed = 1)
R> semE <- searchEM(NULL, heart, stat = "EBIC", search = "global", seed = 1)
R> cc <- droplevels(heart[complete.cases(heart), ])
R> gc <- searchGlobal(stepSearch(minForest(cc), cc), cc, seed = 1)
R> gcEM <- fitModelEM(makeModel(heart, gc$edges), heart)
R> mi <- searchMI(sem, heart, m = 20, search = "global", seed = 1)
\end{CodeInput}
\end{CodeChunk}

\begin{table}[t]
\centering
\begin{tabular}{lrrrr}
\toprule
Procedure & Edges & Dimension & BIC & EBIC \\
\midrule
Complete rows, refitted by EM & 22 & 124 & 44\,179.3 & 44\,298.4 \\
Structural EM, BIC            & 25 & 111 & 43\,801.8 & 43\,937.3 \\
Structural EM, EBIC           & 18 &  82 & 43\,907.3 & 44\,004.7 \\
Multiple imputation           & 24 & 113 & 43\,813.3 & 43\,943.3 \\
\bottomrule
\end{tabular}
\caption{Graphs selected for the heart disease data, with the observed-data criteria of all
920 rows ($\gamma=1/2$ for EBIC).}
\label{tab:heart}
\end{table}

Table~\ref{tab:heart} gives the results. Using all rows lowers the BIC by 377 units with
respect to the graph of the complete rows, and multiple imputation comes within 12 units of
structural EM. In the structural EM graph the diagnosis is adjacent to chest pain, exercise
angina, ST depression, slope, number of vessels, thallium scan, sex, cholesterol and centre;
in the graph of the complete rows it has seven neighbours, among them maximum heart rate but
not cholesterol, sex or angina. Each structural EM search took about three minutes on one
core.

The table also shows a weakness of structural EM. The graph selected under the extended BIC
has a higher observed-data extended BIC, by 67 units, than the graph selected under BIC. The
expected statistics are computed under the current graph, so that the rounds tend to confirm
it (Section~\ref{sec:sem}), and a search can end at a graph that another graph improves on the
observed-data criterion. Comparing the results of several criteria and starting graphs on the
observed-data criterion, as here, reveals such cases.

\subsection{A large data set}
\label{sec:example-large}

The data set ``gene expression cancer RNA-Seq'' of the UCI Machine Learning Repository
\citep{Fiorini:2016} is a random extraction of the TCGA Pan-Cancer RNA-Seq data
\citep{Weinstein+others:2013}: the expression of 20\,531 genes in 801 tumours of five
types. We use it to show the package on a large mixed model. The example is an
illustration of the software, not an analysis or a re-analysis of these gene-expression
data: the two tumour types, the genes and the transformation were chosen for the
illustration. We keep the kidney renal clear cell carcinomas (KIRC, 146 tumours) and the
lung adenocarcinomas (LUAD, 141), and the 13\,719 genes with no zero value among these 287
tumours. Each gene is Box--Cox transformed, with the exponent that maximises the profile
likelihood of a normal model with a mean for each tumour type, so that normality is sought
within types. The heterogeneous BIC forest of all 13\,719 transformed genes and the tumour
type \code{Class} took 133 seconds; the 1521 genes within distance four of \code{Class} in
this forest, standardised, form the data set \code{rnaseq} of \pkg{GraphInfer}, with
1522 variables and 287 observations. This selection keeps genes that are separated from the
tumour type by others, so that the graph on these variables is not trivial.
\begin{CodeChunk}
\begin{CodeInput}
R> data("rnaseq", package = "GraphInfer")
R> f <- minForest(rnaseq, homog = FALSE)
R> s <- stepSearch(f, rnaseq)
R> near <- nearVertices(f, "Class", radius = 2)
R> sub <- rnaseq[, near]
R> g <- searchGlobal(stepSearch(subModel(f, near), sub), sub, rounds = 20, seed = 1)
\end{CodeInput}
\end{CodeChunk}
The forests of the homogeneous and the heterogeneous model took 2.3 seconds each on one
core of an Apple M4 Pro. The heterogeneous forest has a BIC lower by 1722 units, 906\,362.1
against 908\,084.5, and gives \code{Class} 126 neighbours, against 58: the variances of
the genes differ between the two types, which the homogeneous model cannot represent. The
forward search from the heterogeneous forest over all 1522 variables took 3.3 hours
(11\,848 seconds) and added 2179 edges, lowering the BIC to 874\,026.4. In the selected
graph \code{Class} has 310 neighbours, among them all 126 of the forest, and 638, 461 and
112 genes lie at distances two, three and four: these 1211 genes are separated from the
tumour type by their neighbours, although the mean of almost every gene differs between
the types. The largest clique has six variables, the type and five genes, so that each
cell is fitted with about 140 observations, although there are more variables than
tumours.

The neighbourhood of \code{Class} can also be studied on its own
(Section~\ref{sec:local}). On the ball of radius one, the type and its 126 neighbours in
the forest, the forward search took 7.8 seconds and gave 329 edges; the iterated local
search from that graph, with the default 127 perturbation rounds, took 63 minutes and gave
464 edges, lowering the BIC from 67\,737.2 to 67\,302.4. In this ball every gene is a
neighbour of the type in the forest, and all remain so. The ball of radius two has 378
variables; the forward search took 239 seconds and gave 959 edges, with 214 genes adjacent
to \code{Class} and 163 separated from it. Of these 214 neighbours, 197 are neighbours of
\code{Class} in the graph on all variables as well. The iterated local search with 20 rounds
took 3.1 hours (11\,072 seconds) and gave 1078 edges, lowering the BIC from 199\,341.2 to
199\,160.8; \code{Class} then has 223 neighbours and 154 genes are separated from it. Of
the 203 neighbours of \code{Class} in the graph on all variables that lie in this ball,
193 are among these 223.
The local graphs describe the marginal distribution of the variables of the ball
(Section~\ref{sec:local}): an agreement with the graph on all variables is reassuring, but
it is a check rather than a proof. The times grow quickly with the number of variables, from
7.8 seconds for 127 to 239 for 378 and 11\,848 for 1522 in the forward search, and a global
search over all 1522 variables is out of reach in pure \proglang{R}
(Section~\ref{sec:study-global}).

\subsection{Accumulation trees}
\label{sec:acctree-example}

The last example uses simulated data, so that the truth is known. We simulate 500
lineages from a random tree with 12 events, flipping each recorded event with
probability 0.02:
\begin{CodeChunk}
\begin{CodeInput}
R> tr <- rAccTree(12, seed = 3)
R> D <- rAccData(tr, 500, noise = 0.02, format = "factor", seed = 3)
R> gComp <- searchGlobal(NULL, D, compatible = TRUE, seed = 1)
R> gFree <- searchGlobal(minForest(D), D, seed = 1)
R> sapply(list(compatible = gComp, unrestricted = gFree,
+              truth = inducedGraph(tr)), function(g) BIC(fitModel(g, D)))
\end{CodeInput}
\begin{CodeOutput}
  compatible unrestricted        truth
    2309.569     2288.037     2356.542
\end{CodeOutput}
\end{CodeChunk}
The noise shows at once. The unrestricted search leaves the compatible class, and the
induced graph of the true tree has the worst BIC of the three, as is to be expected
when noisy events no longer screen exactly. The full accumulation model of
\fct{accFit} makes the noise explicit:
\begin{CodeChunk}
\begin{CodeInput}
R> acc <- accFit(D)
R> compareTrees(acc$tree, tr)
\end{CodeInput}
\begin{CodeOutput}
  events differ relative
1     12      0        0
\end{CodeOutput}
\end{CodeChunk}
The event tree is recovered without error in its undirected structure: no event has a
different split in the two trees. The two levels of \fct{accTest} then answer different questions:
\begin{CodeChunk}
\begin{CodeInput}
R> accTest(D, B = 99, level = "graph", seed = 1)
R> accTest(D, B = 99, level = "full", fdb = TRUE, seed = 1)
\end{CodeInput}
\end{CodeChunk}
At the graph level, which assumes noise-free events, the accumulation hypothesis is
rejected: the difference in BIC is 27.0 with bootstrap p-value 0.01. At the full level,
the difference in BIC between the accumulation model and the selected graph is 34.1,
with bootstrap p-value 0.12 and fast double bootstrap p-value 0.13, and the hypothesis
is not rejected. Since these data do come from an accumulation tree, the full level
gives the correct conclusion here, and the rejection at the graph level reflects the
noise rather than a failure of the mechanism. With 12 events and $n=500$ the fitted accumulation model is well determined, the situation
in which the calibration studies showed no substantial problem
(Section~\ref{sec:study-acctest}); one simulated data set illustrates the behaviour but does
not establish the calibration.

\section{Numerical assessment}
\label{sec:numerical}

This section reports simulation studies of the statistical behaviour of the package and
of the cost of its searches. Each subsection answers one question, and the practical
conclusions motivate the defaults. The data are simulated, so that the truth is known.
The studies of the global and compatible searches (Sections~\ref{sec:study-global}
and~\ref{sec:study-compatible}, except the tree-move comparison) were run with a
development version of 2 October 2026; we repeated 19 of their runs with the version
described here, with the same models and seeds, and obtained identical BIC values, so that
their results stand, and their times refer to the original runs. The other results were
computed with the version described here. The full tables are in the vignettes
\code{GraphInfer-global}, \code{GraphInfer-missing} and \code{GraphInfer-exact} of the
package, and the replication material reproduces every number.

\subsection{Accelerated and reference stepwise search}
\label{sec:benchmark-step}

Section~\ref{sec:stepwise} states that the fast engine reorganises the greedy calculation
and does not approximate it. We simulated continuous data in which the $p$ variables fall
into blocks, each driven by one latent variable, and ran the forward search from the
forest with both engines (Table~\ref{tab:engines}). Wherever both were run, they added the
same edges in the same order. The ratio of their times grew from about 35 at $p=20$ to
1060 at $p=80$, since the classic engine scans all candidate non-edges at each step. The
fast engine reached $p=1200$ in 22 minutes; its time grew about as $p^{3.5}$ over the
range of the table, in which the number of edges per variable also rose, from 2 to 6. The forest is a small part of this cost, 0.84 seconds at $p=1200$.

For mixed data the forest statistics used to be computed pair by pair, one call of the
likelihood core per pair; they are now computed for all pairs of a kind at once
(Appendix~\ref{app:structure}).
On the gene-expression data of Section~\ref{sec:example-large}, with 1521 genes, the
tumour type and 287 observations, the statistics of the 1\,157\,481 pairs took about 43
minutes pair by pair and 0.3 seconds vectorised, for the homogeneous and the heterogeneous
model alike, with relative differences below $6\times10^{-13}$
(Table~\ref{tab:forest-mixed}); the whole forest now takes 2.3 seconds, most of it in
Kruskal's algorithm. The statistics are the same; only their computation changed, within
pure \proglang{R}.

\begin{table}[t]
\centering
\begin{tabular}{rlrrrr}
\toprule
Genes & Model & Pairs & Pair by pair (s) & Vectorised (s) & Ratio \\
\midrule
200 & homogeneous   &      20\,100 &    1.48 & 0.006 &  246 \\
     & heterogeneous &      20\,100 &    1.46 & 0.006 &  243 \\
1521 & homogeneous   & 1\,157\,481 & 2571.6  & 0.30  & 8659 \\
     & heterogeneous & 1\,157\,481 & 2492.5  & 0.29  & 8536 \\
\bottomrule
\end{tabular}
\caption{Pairwise statistics of \fct{minForest} for the gene-expression data of
Section~\ref{sec:example-large} (the first 200 genes, or all of them, with the tumour type),
in one run on one core of an Apple M4 Pro.}
\label{tab:forest-mixed}
\end{table}

\begin{table}[t]
\centering
\begin{tabular}{rrrrr}
\toprule
$p$ & $n$ & Edges selected & Classic engine (s) & Fast engine (s) \\
\midrule
  20 &  400 &   39 &  0.27 &    0.01 \\
  35 &  500 &   84 &  3.4  &    0.03 \\
  50 &  600 &  133 & 15.7  &    0.06 \\
  80 &  800 &  270 & 157.4 &    0.15 \\
 200 & 1000 &  828 &       &    2.0  \\
 500 & 1500 & 2611 &       &   55.8  \\
1200 & 2000 & 7450 &       & 1349.4  \\
\bottomrule
\end{tabular}
\caption{Forward search from the forest on simulated continuous data, on one core of
an Apple M4 Pro. The classic engine was run up to $p=80$.}
\label{tab:engines}
\end{table}

\subsection{Global search}
\label{sec:study-global}

How much effort does \fct{searchGlobal} need, and which default follows? We simulated
dense Gaussian models with \fct{rModel}, with 3.5 to 3.9 true edges per variable,
$n=1000$, 20 models per setting and several seeds per model, all started from the forest.
A run is said to reach the truth when its BIC is at least as good as that of the generating
graph; this concerns the criterion, not the recovery of the true graph. With $p=30$,
iterated local search with a kick of 25 edges and 50 rounds reached the truth in 90\% of
the runs, against 75\% for the scheduled search at about five times the cost and 80\% for
annealing at about 24 times; at $p=50$ and 80 annealing and the scheduled search did much
worse than ILS at 15 to 25 times its cost. Table~\ref{tab:kick} shows how the effort must
grow with the problem: a kick of a quarter to a third of the true edges gave 82--90\% at
every $p$, which motivates the default of a quarter of the edges of the first local
optimum. Rounds matter at least as much: with a kick of 100 but 20 instead of 50 rounds,
69\%, 47\% and 18\% of the runs reached the truth at $p=50$, 80 and 120. A second study,
with the kick as a fraction of the first local optimum, found sparse problems easy and
dense ones better served by more rounds than by larger kicks: at $p=120$, a fraction of
0.25 with 100 rounds reached the best value found in 50\% of the runs, against 10\% for a
fraction of 0.5 with 50 rounds, at equal cost. Hence the default fraction 0.25 and
$\max(50,p)$ rounds. For large dense problems the optimum is hard to find, and different
runs end at different values; one should run several seeds, read the trace of the search,
and increase the number of rounds while the best value still improves. With $p=120$ and
dense models a run of 100 rounds took a median of 40 minutes, which is why the global
search over all variables of Section~\ref{sec:example-large} is out of reach.

\begin{table}[t]
\centering
\begin{tabular}{rrrrrr}
\toprule
$p$ & True edges & Kick 10 & Kick 25 & Kick 50 & Kick 100 \\
\midrule
 30 & 106 & 76\% & 90\% & 92\% & 93\% \\
 50 & 190 & 36\% & 74\% & 88\% & 94\% \\
 80 & 304 &  9\% & 46\% & 76\% & 88\% \\
120 & 464 &  0\% &  8\% & 45\% & 82\% \\
\bottomrule
\end{tabular}
\caption{Percentage of runs of iterated local search reaching the truth, with 50 rounds.
True edges: median over the models.}
\label{tab:kick}
\end{table}

\subsection{Search within the accumulation-tree compatible class}
\label{sec:study-compatible}

How hard is the compatible class of Section~\ref{sec:acctree-search}, and do the tree moves
overcome the obstruction? For six and seven events all decomposable graphs can be
enumerated (18\,154 and 617\,675, of which 5\,003 and 68\,399 are compatible), so that the
best graph of each class is known exactly. We used data from 400 simulated accumulation
trees of four shapes, with $n=200$ and 1000. Table~\ref{tab:compat-exact} shows that the
compatible class, although small, is rugged: with single-edge moves almost one graph in
five is a local optimum, and steepest descent reaches the best graph from fewer than one
graph in a hundred. Swap moves reduce the number of local optima, but, as
Corollary~\ref{cor:unreachable} states, neither they nor single-edge moves reach a graph
with a clique of four or more events, so that the best graph of star trees is never found.
The global search with tree moves reached the exact optimum of the compatible class in all
2000 runs (400 data sets, five seeds each; median 0.3 seconds per run). This is evidence
about the search on these data, not a guarantee of optimality.

\begin{table}[t]
\centering
\small
\setlength{\tabcolsep}{4pt}
\begin{tabular}{lrrrr}
\toprule
 & All decomposable & \multicolumn{3}{c}{Compatible} \\
\cmidrule(lr){3-5}
 & graphs & edge moves & edge moves and swaps & tree moves \\
\midrule
Graphs in the class         & 617\,675 & 68\,399 & 68\,399 & 68\,399 \\
Local optima (\%)           & 0.0--1.5 & 16--21  & 2       & \\
Descent reaches best (\%)   & 62--78   & 0.2--0.7 & 28--60 & \\
Same, star trees (\%)       & 100      & 0       & 0       & \\
Search reaches best (\%) & 100  & 37--100 & 65--100 & 100 \\
Same, star trees (\%)       & 100      & 0       & 0       & 100 \\
\bottomrule
\end{tabular}
\caption{Seven events: local optima and steepest descent by complete enumeration, and runs
of the global search (five seeds per data set) reaching the exact optimum of the class.
Medians over data sets; ranges over tree shapes, with star trees apart.}
\label{tab:compat-exact}
\end{table}

For 15 and 30 events the optimum is not known. With 10 trees per shape, $n=1000$ and three
runs each, every run with tree moves reached the best compatible BIC found, which equalled
that of the true graph in every case; the true graph's BIC is an upper bound for the
minimum, so that this shows that the runs agreed with each other and with the truth. Runs
with edge moves did so only for chains, and ended a median of 97, 336 and 1\,170 BIC units
above for binary, random and star trees. An unrestricted search reaches the same BIC on
such data, except for star trees with 30 events, but its result is often not compatible,
because several graphs fit equally well. To select a compatible model one should therefore
use \code{compatible = TRUE} with the default tree moves.

\subsection{Level of the exact tests}
\label{sec:study-exact}

Section~\ref{sec:exact} claims exactness in finite samples, and the question is how
far the chi-squared approximation is from it. We simulated from a mixed model
on two discrete variables, with three and two levels, and three continuous ones,
without two of the edges, and tested it against the complete graph 4000 times with
\fct{lrtGauss}. In the heterogeneous design, 2902 of the 4000 data sets could be tested;
the others had too few observations in a cell, which depends on the discrete data only,
and the test is exact conditionally on them. For edges between discrete variables we
simulated from a model on three discrete variables, with three, two and two levels, and two
continuous ones, whose larger graph has the discrete triangle and the clique of the first
discrete variable and the two continuous ones: once without the two edges of the third
discrete variable, and once without an edge between discrete variables and an edge between a
discrete and a continuous variable, tested jointly (4000 data sets each, 1000 conditional
simulations per test). Table~\ref{tab:exact-level} shows that the
exact test holds its level, whereas the chi-squared approximation rejects a true
model several times too often, most of all in the heterogeneous design, in which each
cell estimates its own variances, and also when discrete edges are tested in small
samples.

\begin{table}[t]
\centering
\begin{tabular}{lrrrrr}
\toprule
 & & \multicolumn{2}{c}{Level 0.05} & \multicolumn{2}{c}{Level 0.01} \\
\cmidrule(lr){3-4}\cmidrule(lr){5-6}
Design & Replicates & Exact & Chi-squared & Exact & Chi-squared \\
\midrule
Homogeneous, $n=25$   & 4000 & 0.051 & 0.144 & 0.010 & 0.048 \\
Heterogeneous, $n=60$ & 2902 & 0.048 & 0.384 & 0.008 & 0.206 \\
\addlinespace
Discrete, $n=40$ & 4000 & 0.049 & 0.164 & 0.011 & 0.060 \\
Both kinds, hom., $n=40$ & 4000 & 0.055 & 0.164 & 0.013 & 0.059 \\
Both kinds, het., $n=60$ & 4000 & 0.049 & 0.114 & 0.010 & 0.031 \\
\bottomrule
\end{tabular}
\caption{Rejection rates under the null hypothesis. First two rows: two edges with a
continuous endpoint; last three (conditional test): two edges between discrete
variables, or one between discrete variables and one with a continuous endpoint (``both
kinds''), in a homogeneous or heterogeneous model. The standard error of a rate near 0.05 is
about 0.004, and near 0.01 about 0.0016.}
\label{tab:exact-level}
\end{table}

\subsection{Calibration of the accumulation tests}
\label{sec:study-acctest}

How reliable are the bootstrap p-values of \fct{accTest}? At the full level, with eight
events, $n=300$, $B=99$ and 20 data sets per setting generated by accumulation trees, the
criterion statistic rejected 1, 0 and 0 data sets at the 5\% level without noise, with 3\%
noise, and with 3\% noise and 10\% missing values; in these configurations the fitted model
is well determined. With 20 data sets, this shows no substantial calibration problem, but
cannot establish the level precisely.
The bootstrap becomes liberal when many estimates lie on the boundary, because the simulated
data are then sparser than data from the true model. We simulated from a tree with 98
parameters, fitted to a real data set of 105 units that will be reported in a later version,
with $n=105$, 48 data sets and $B=19$; the estimated dimension averaged 83 on these data sets
and 75 on their bootstrap data sets. Table~\ref{tab:acctest-cal} shows that the bootstrap
p-values were too small, that a bootstrap from the true model gave p-values compatible with
uniformity, which places the error in the plug-in estimates and not in the statistic or the
searches, and that the fast double bootstrap removed most of the error for the criterion,
less for the likelihood contrast. In a still sparser configuration, 57 events and $n=21$,
the plain bootstrap rejected 7 of 20 null data sets at the 5\% level. These figures concern few configurations and replicates: they show no substantial
calibration problem when the fitted model is well determined, and the test should not be
regarded as calibrated in general. At the graph level, which assumes
events recorded without error, 9 of 20 data sets with eight events, $n=500$ and 5\% noise
were rejected, as Table~\ref{tab:acctest-levels} anticipates.

\begin{table}[t]
\centering
\small
\begin{tabular}{lrrr}
\toprule
Bootstrap ($n=105$, 48 data sets, $B=19$) & Mean p-value & At most 0.05 & Uniformity \\
\midrule
Plain, criterion                       & 0.31 & 10 & \\
Fast double, criterion                 & 0.47 & 4  & $p=0.20$ \\
Fast double, likelihood contrast       & 0.43 & 4  & $p=0.035$ \\
From the true model, criterion         & 0.55 & 2  & \\
\bottomrule
\end{tabular}
\caption{Full level of \fct{accTest} under the null hypothesis. Uniform p-values on this grid
have mean 0.525; ``Uniformity'': Cram\'er--von Mises test.}
\label{tab:acctest-cal}
\end{table}

\subsection{Missing-data inference and structure inference}
\label{sec:study-missing}

What is lost by missing values, and which procedure should be the default? We simulated
homogeneous models from \fct{rModel} with about three edges per variable: continuous
($p=10$, 30), binary ($p=10$, 16) and mixed (five binary and five continuous variables, or
six binary and 24 continuous), keeping only models in which every discrete level has
probability at least 0.1. We used $n=200$ and 1000, with 10\% or 30\% of all values missing,
missing completely at random (MCAR), at random (MAR: half of the variables always observed,
the others missing with a probability that increases, on the logistic scale, with the
observed variable most correlated with them) or not at random (MNAR: a probability
increasing with the value itself). There were 20 data sets per setting, 1440 in all, and
selection used BIC throughout.

For estimation with the true graph, measured by the Kullback--Leibler divergence of the
fit from the truth, EM lost little under MCAR and MAR (median 0.027 to 0.12 with $n=1000$,
against 0.022 with all values known). Single imputation was 2 to 7 times worse, a gap that
does not close with $n$ since imputed values have too little variability. Complete rows
were worst, and often unusable: with 30\% missing and $p=30$ they gave no fit at all,
whereas EM always gave one. Under MNAR all methods are biased, and more data do not help.

Table~\ref{tab:miss-select} concerns selection with 30\% missing values. Each structural
EM search started from the forest of the data with missing values replaced by the mean or
the most frequent level; single imputation used \fct{imputeMissing} under the graph of a
single-stage search, and multiple imputation used \fct{searchMI} with 20 imputations under
the denser graph of a structural EM with AIC. Complete rows lose most of the information
and add false edges, and a selection was possible in only 45\% of all the data sets. A single
stage of structural EM is conservative, as Section~\ref{sec:sem} leads one to expect; the
two-stage default finds as many true edges as multiple imputation. With 10\% missing all
procedures but complete rows were within three points of the reference, and the two stages
found 86--88\% of the true edges against 83--84\% for the reference, a stepwise search
under BIC alone, at a small cost in false edges. Within a single stage, the global search
was better than the stepwise one in 41\% of the data sets, by a median of 30 BIC units, and
worse in 0.5\%.

\begin{table}[t]
\centering
\small
\setlength{\tabcolsep}{4pt}
\begin{tabular}{lrrrrrr}
\toprule
 & All values & Complete & Single & \multicolumn{2}{c}{Structural EM} & Multiple \\
\cmidrule(lr){5-6}
Mechanism & known & rows & imputation & One stage & Two stages & imputation \\
\midrule
MCAR & 84 / 2 & 45 / 22 & 76 / 7  & 76 / 2 & 82 / 7  & 80 / 6 \\
MAR  & 85 / 2 & 39 / 18 & 71 / 18 & 66 / 7 & 74 / 8  & 73 / 8 \\
MNAR & 84 / 2 & 31 / 24 & 65 / 19 & 62 / 9 & 73 / 16 & 73 / 20 \\
\bottomrule
\end{tabular}
\caption{Selection with 30\% missing values and $n=1000$: true edges found and false edges, as
percentages of the true edges, averaged over all model types. Structural EM with global
search. Complete rows: only the data sets where a selection was possible.}
\label{tab:miss-select}
\end{table}

Table~\ref{tab:miss-stages} compares the procedures on the criterion they all aim at, the
observed-data BIC, in 240 of the data sets, with 30\% missing under MCAR or MAR. Paired by
data set, two stages reached a lower BIC than one in 73\% of the data sets with stepwise
rounds and in 63\% with a final global search, and a higher one in 15\% and 21\%, at 2.2 and
1.2 times the median cost. Two stages with a global search reached the lowest value most
often, and beat multiple imputation in 35\% of the data sets against 17\% the other way, at a
lower median cost. With much missing data we therefore recommend
\code{searchEM(..., search = "global")}, which runs two stages by default; the stepwise
default costs less and is reasonable for a first analysis.

\begin{table}[t]
\centering
\begin{tabular}{lrrrr}
\toprule
 & Structural EM, & Structural EM, & Structural EM, & Multiple \\
Model & one stage, global & two stages, stepwise & two stages, global & imputation \\
\midrule
Continuous & 31 (6.5)  & 60 (1.4)  & 74 (5.7)  & 59 (6.6)  \\
Binary     & 34 (14.1) & 59 (11.3) & 64 (18.0) & 57 (65.7) \\
Mixed      & 32 (28.3) & 39 (7.3)  & 68 (22.6) & 45 (33.3) \\
\bottomrule
\end{tabular}
\caption{Percentage of data sets in which each procedure reached the lowest
observed-data BIC, ties counting for each, and in parentheses the median seconds per
data set. 30\% missing under MCAR or MAR, 240 data sets; a single stepwise stage, also
compared, reached it in 14--31\%.}
\label{tab:miss-stages}
\end{table}

Many true edges of the discrete and mixed models are too weak to be found at these sample
sizes. We call a true edge $\{u,v\}$ detectable when $2nI>\log n$, where $I$ is the
conditional mutual information of $u$ and $v$ given all other variables under the true
distribution, so that $2nI$ is the expected likelihood-ratio statistic of the edge. With
$n=200$ and 1000, 50\% and 29\% of the true edges were not detectable in the binary models,
26\% and 15\% in the mixed and 9\% and 0\% in the continuous ones. With all values known BIC
found 91\% of the detectable edges; the two-stage global search found 90--94\% of them with
10\% missing and 69--84\% with 30\%, and with \code{stat = "EBIC"} 3 to 9 points fewer, with
half as many false edges. Finally, the conditional mean under the graph selected by
structural EM predicted missing continuous values nearly as well as under the true graph
(root mean squared error 0.85 against 0.83 standard deviations under MCAR, 1.00 for the
observed mean), whereas a random draw gave 1.19, as it should: draws are for multiple
imputation, not for prediction.

\section{Discussion}
\label{sec:discussion}

\pkg{GraphInfer} rests on one likelihood calculation and one definition of model dimension,
used for Gaussian, discrete and mixed data, by the forest, stepwise and global searches, EM
and structural EM. Results of different procedures can therefore be compared directly: in
Section~\ref{sec:workflow-example} homogeneous and heterogeneous models and different search
routes, in Section~\ref{sec:missing-example} a graph selected on the complete rows,
structural EM and multiple imputation, all on the same observed-data criterion. With empty
cells the dimension is an adaptive one, based on the observed support, and tests use
degrees of freedom of their own (Section~\ref{sec:likelihood}).

This common calculation depends on decomposability, which is a real restriction. A graph
with a chordless cycle cannot be fitted, and strong decomposability excludes further graphs
in the mixed case. The result of a search is the best decomposable graph found, and it may contain an edge with a very small estimated partial association that cannot be
deleted on its own without destroying decomposability, as Murder--Frost in
Section~\ref{sec:example-gauss}. A missing edge can be read as a conditional independence
under the selected model, whereas a present edge is weaker evidence of direct dependence.
The wider class of \pkg{gRim} is the alternative when this matters, at the price of
iterative fitting.

The searches of Section~\ref{sec:using} show that the effort matters: on the \code{wine}
data the greedy searches stopped up to 181 BIC units above the best value found. A
stochastic search gives no certificate of optimality, and for large dense problems
different runs end at different values. Selection also affects what may be concluded
afterwards. The p-values of \fct{drop1} and \fct{lrtGauss} are valid for a comparison fixed
in advance; applied to a graph selected on the same data they take the graph as given and
do not account for its selection, and the exactness shown in Section~\ref{sec:study-exact}
concerns fixed hypotheses only. The edge frequencies of \fct{searchMI} measure stability
across imputed data sets, not the sampling variability of the selected graph. The methods
for incomplete data assume that the values are missing at random; differences between
complete and incomplete rows, as in the coronary artery disease data, argue against
discarding the incomplete rows but cannot confirm this assumption.

Accumulation trees show how a substantive mechanism becomes a restricted model space within
the framework. The likelihood of a graph is unchanged, but two things had to be added:
moves that respect the class, because single-edge moves cannot reach part of it
(Corollary~\ref{cor:unreachable}), and a full model for what the graph cannot express, the
support restrictions and the measurement noise. The two levels of \fct{accTest} can give
different answers, and the bootstrap of the full level showed no substantial calibration
problem only in configurations where the fitted model is well determined.

Writing the package in pure \proglang{R} keeps every calculation readable and makes it a
reference against which faster implementations can be checked. Exploiting the structure of
the calculations goes a long way within \proglang{R}: the fast stepwise engine follows the
same path as the exhaustive one at up to a thousandth of its cost, and the vectorised
forest statistics for mixed data took 0.11 seconds instead of 414 for 1001 variables; the cost is in the size of
the problems that can be treated, which is the purpose of the planned companion
\pkg{GraphInferComplex}. Several limitations of the present version are open problems.
Exact tests are not available for data with missing values, nor in the mixed case with
case weights or a covariate design. EM for mixed data is limited to
$10^5$ discrete cells, a limit that also binds structural EM. The accumulation model
requires binary events, its tree search is local, and tree moves require data that are all
discrete or all continuous. Finally, a hypothesis stated as a non-decomposable graph must at
present be embedded in a decomposable one by hand.

\section*{Computational details}

All results were computed with \proglang{R} 4.5.0 on macOS 15.7 with an Apple M4 Pro
processor, with \pkg{GraphInfer} 0.1.0; times refer to one core.
The data sets \code{rnaseq} (Sections~\ref{sec:example-large}
and~\ref{sec:benchmark-step}), \code{heartdisease} and \code{wine} of \pkg{GraphInfer} were prepared
from the files of the UCI Machine Learning Repository (doi:10.24432/C5R88H,
doi:10.24432/C52P4X and doi:10.24432/C5PC7J, all licensed CC BY 4.0) by scripts of the
replication material; \code{wine} coincides with the data set of the same name in
\pkg{gRbase}. The replication material contains one script for each
example and each study, the stored results of the simulation studies, and their outputs;
the seeds are set in the scripts, and the replicates of a study draw their own seeds, so
that the results do not depend on the order or the parallelism of the runs.
\pkg{GraphInfer} 0.1.0 will be submitted to the Comprehensive \proglang{R} Archive
Network (CRAN); until it is available there, the package and the replication material
are available from the author on request. A later version of this manuscript will
refer to the CRAN release.

\section*{Acknowledgements}

The author thanks the Funda\c{c}\~ao A.~Plaus\^onio and RLstatLab for partial support.

\appendix

\section{Architecture and implementation}
\label{app:structure}

\pkg{GraphInfer} has 48 exported functions and about 9\,800 lines of \proglang{R},
documentation comments included.
The code is organised in thirteen modules, which can be grouped in four layers. The
base consists of the model object (the class, \fct{makeModel} and the methods), the
graph utilities (perfect orderings, clique trees, and the tests of decomposability and
strong decomposability) and the likelihood core. The core rests on two internal
functions: \code{.margLik} gives the saturated marginal log-likelihood and dimension of
a vertex set, and \code{.edgeStat} the four-term difference that gives the statistic
of an edge. \fct{fitModel} adds up the marginal terms over cliques and separators.

The second layer contains the structure-inference modules: forests, stepwise search,
with its fast and classic engines, global search, and local analysis. These modules
propose graphs and obtain their deviance differences from the core. The only
exception is that \fct{minForest} computes the pairwise statistics of all pairs at once,
for continuous, discrete and mixed data, by vectorised formulas, which the tests compare
with the pair-by-pair calculation of the core. The third layer contains the exact tests
and the covariate adjustment.

The fourth layer builds on the others. The missing-data module has its own E-steps and
M-steps, because the observed-data likelihood does not factorise over cliques, but the
structural EM reuses the searches unchanged by handing them a weighted pseudo-sample
(Section~\ref{sec:sem}). The accumulation-tree module reuses the same score with a
compatibility check on each move, adds the tree moves, and has a separate likelihood
for the latent-class model of \fct{accFit}. Two support modules provide
simulation, and the running of replicates, which gives each replicate its own seed so
that the results do not depend on how they are run.

The complete tables of functions and calls are in the directory
\code{extdata/structure} of the installed package, and the vignette
\code{GraphInfer-structure} draws the workflow and the call graph from them. A test
keeps these tables in agreement with the code.

\paragraph{Implementation details.}
The fast forward engine associates the score of each add-eligible non-edge with its
separator, caches it, and recomputes it only when an addition changes the neighbourhood of
its endpoints; admissibility is checked locally, without testing the chordality of the whole
proposed graph. \fct{minForest} computes the statistics of all pairs at once: continuous
pairs from the cross-products of the centred data, discrete pairs from the cross-products of
a one-hot coding of the levels, and each discrete variable against all continuous ones from
their within-cell residuals; the tests compare these with the pair-by-pair four-term
calculation of the likelihood core. For discrete data with more than $10^5$ cells, EM holds
the clique and separator tables and lists, for each distinct incomplete row, the
completions of its missing values with their posterior probabilities; these weighted
completions are the expected complete data, so that the iterates are those of the full-table
EM, at a cost set by the number of missing values per row. A row with more than 65\,536
completions (17 or more missing binary values) is treated by propagation on the junction
tree \citep{Lauritzen:1995}, and a clique or separator with more than $10^7$ cells is held
only on the cells reached by some completion; the listed completions are limited to
$2\times10^6$ in all. The options \code{GraphInfer.discreteEM},
\code{GraphInfer.maxRowCompletions} and \code{GraphInfer.maxCompletions} control these
choices. The boundary refit of Section~\ref{sec:em} fixes at zero the cells of the clique
tables with expected count at most $10^{-3}$ and keeps the result if the observed-data
log-likelihood drops by at most $10^{-6}$; if the cells fail together, and there are at most
20 of them, they are tried one at a time (option \code{GraphInfer.emBoundary}). The global
searches track the criterion incrementally after each move, and the tree moves recompute
only the clique and separator terms of the nodes they change.

\section{Validation}
\label{app:validation}

The package is validated by comparison with references that do not share its code, rather
than against stored output of earlier versions. Table~\ref{tab:validation} groups the tests
by the kind of reference. The test suite of the version described contains
13\,724 expectations in 52 files, all of which pass; 5\,830 of them are one
enumeration of local compatibility checks, on all compatible graphs with five vertices. Files that take more than about a second are
skipped on CRAN and run by \code{devtools::test()} and \code{devtools::check()}. Recent
additions compare the closed-form test degrees of freedom of Section~\ref{sec:fitting} with
the ranks of the parameter designs on every single-edge move of sparse discrete and mixed
data, and the vectorised forest statistics for mixed data with the pair-by-pair
calculation.

\begin{table}[t]
\centering
\small
\begin{tabular}{p{30mm}p{109mm}}
\toprule
Reference & What is compared \\
\midrule
Closed forms &
Gaussian and homogeneous mixed likelihoods with the chain rule of regressions,
heterogeneous likelihoods with per-cell fits, edge statistics with differences of fitted
models, one-edge exact p-values with the Beta tail, EM under monotone missingness with
the factorised maximum-likelihood estimate, and the accumulation model without noise
with its closed form. \\
\addlinespace
Independent \proglang{R} fits &
\fct{lm} for likelihoods, added edges, residualisation and the one-edge exact tests,
which are its $F$ tests; \fct{loglin} for discrete fits, added edges and collapsibility. \\
\addlinespace
Complete enumeration &
The global search and the tree moves against the best of all decomposable or compatible
graphs with five vertices, under BIC and the extended BIC; the forest searches, with case
weights and with missing values, against all forests; decomposability against simplicial
elimination and forbidden paths; separation against the transitive closure. \\
\addlinespace
Numerical optimisation &
EM with missing discrete values, and \fct{accFit} for a given tree, against numerical
maximisation; a clique of $2^{26}$ cells against the optimality conditions; the
observed-data log-likelihood never decreases along EM. \\
\addlinespace
Equivalent engines &
Fast and classic stepwise engines, which select the same graph on continuous, discrete
and mixed data; full-table, row-wise and sparse EM; propagated and listed rows; the
incrementally tracked criterion of the global search and of the tree moves against a
full refit; local and full compatibility checks; integer case weights against repeated
rows. \\
\addlinespace
Distributions &
Uniform p-values of exact tests along removal chains under the smaller model; the Monte
Carlo heterogeneous test against the Beta law when there is one cell; imputation draws
against the closed-form conditional distributions; \fct{rModelData} against the true
distribution; fast double bootstrap p-values against hand-computed cases. \\
\addlinespace
\raggedright Interfaces and invariance &
The argument order and defaults of \pkg{gRapHD} in \fct{minForest} and \fct{stepw};
invariance of edge statistics to the units of continuous variables; replicates that
give the same result in any order; call tables that list exactly the functions of the
package. \\
\bottomrule
\end{tabular}
\caption{Validation by type of reference.}
\label{tab:validation}
\end{table}

With \pkg{gRapHD} 0.2.10, clique decompositions, the deviances of fitted continuous,
discrete and mixed models, edge deviances on dense data and the maximum-likelihood forest of
a continuous example all agreed, in 25 checks; dimensions differ by design, because
\pkg{gRapHD} does not impose the sum-to-one constraint on cell probabilities. With \pkg{gRim}
0.3.4, the AIC and BIC of the graph selected by its forward search on the \code{wine} data,
6631.612 and 6806.610, were identical to those of \pkg{GraphInfer} for the same homogeneous
model.

\section{Exact tests in mixed models: proofs}
\label{app:exactproofs}

Let $G$ be strongly decomposable and let $\{u,v\}$, with at least one continuous endpoint,
lie in a unique maximal clique $C$ of $G$ and of its star-augmented graph, so that
$G'=G-\{u,v\}$ is strongly decomposable; let $S=C\setminus\{u,v\}$.

\begin{lemma}\label{lem:side}
Let $A_u$ and $A_v$ be the components of $G'\setminus S$ that contain $u$ and $v$. Then
$A_u\neq A_v$, and at least one of them contains no discrete vertex; if $v$ is discrete,
$A_u$ contains none.
\end{lemma}

\begin{proof}
Suppose that $u$ and $v$ are joined in $G'$ by a path avoiding $S$, and take a shortest
one, $u=x_0,x_1,\ldots,x_r=v$. If $r=2$, then $\{u,v,x_1\}$ is a triangle of $G$ and lies in
a maximal clique that contains $x_1\notin C$, so that $\{u,v\}$ lies in two maximal cliques.
If $r\ge3$, the path and the edge $\{u,v\}$ form a cycle of length at least four in $G$; a
chord would join two non-consecutive vertices of the path, contradicting its minimality, so
the cycle is chordless, contradicting chordality. Hence $A_u\ne A_v$. If both contained a
discrete vertex, let $\delta_u$ be one nearest to $u$ in $A_u$ and $\delta_v$ one nearest to
$v$ in $A_v$ (with $\delta_v=v$ if $v$ is discrete). The shortest paths to them have
continuous interiors, and joined through the edge $\{u,v\}$ they form a path in $G$ between
two discrete vertices that are not adjacent, since they lie in different components of
$G'\setminus S$ and differ from $\{u,v\}$; its interior is continuous, so it is a forbidden
path, contradicting strong decomposability. If $v$ is discrete it lies in $A_v$, so $A_u$
contains no discrete vertex.
\end{proof}

\begin{proposition}\label{prop:onestep}
Let $w$ be an endpoint whose component $A_w$ is continuous (Lemma~\ref{lem:side}) and $z$ the
other endpoint. Let $D$ be the step deviance $D_{uv\mid S}$ of \eqref{eq:fourterm}.
\begin{enumerate}
\item[(a)] In the homogeneous model $\exp(-D/n)=\mathrm{RSS}_1/\mathrm{RSS}_0$, the ratio of the
residual sums of squares of the least-squares regressions of $Y_w$ on $X_1$ and on $X_0$,
where $X_0$ contains the indicators of the cells of $\Delta_S$ and $Y_{\Gamma_S}$, and $X_1$
adds $Y_z$ if $z$ is continuous, or replaces the cells of $\Delta_S$ by those of
$\Delta_S\cup\{z\}$ if $z$ is discrete. In the heterogeneous model $D$ is the expression of
Section~\ref{sec:exact}, with the regressions fitted within the cells of $\Delta_S$.
\item[(b)] Under $G'$, conditionally on the discrete observations and on the continuous
variables outside $A_w$, the rows of $Y_w$ are independent with
$Y_w\sim N\bigl(a(i_{\Delta_S})+b^\top y_{\Gamma_S},\sigma^2\bigr)$, where $b$ and $\sigma^2$ also
depend on $i_{\Delta_S}$ in the heterogeneous model. Consequently, if $X_1$ has full rank $r_1$
and $X_0$ rank $r_0$, then $\exp(-D/n)\sim\operatorname{Beta}\bigl((n-r_1)/2,(r_1-r_0)/2\bigr)$
in the homogeneous model, and in the heterogeneous model the residual sums of squares within
groups and the extra sums of squares are independent $\sigma_j^2\chi^2$ variables with the
degrees of freedom of Section~\ref{sec:exact}.
\end{enumerate}
\end{proposition}

\begin{proof}
(a) By Corollary~1 of \citet{Frydenberg+Lauritzen:1989}, $D$ is the likelihood-ratio
statistic for $u\perp v\mid S$ in the saturated marginal model on $C$, which is
\eqref{eq:fourterm}. In both cases the discrete parts of the four terms cancel: for a continuous $z$ the four
sets have the same discrete part, and for a discrete $z$, $\Delta_{Suv}=\Delta_{Sz}$ and
$\Delta_{Sw}=\Delta_S$. For a set $A$ and a continuous $w\notin A$, $|W_{A\cup\{w\}}|/|W_A|$ is
the Schur complement of $W_A$, the residual variance of $Y_w$ after regression on
$Y_{\Gamma_A}$ with the pooled within-cell scatter, that is $\mathrm{RSS}/n$ for the regression
on the cell indicators of $\Delta_A$ and $Y_{\Gamma_A}$. The four terms therefore give
$D=n\log(\mathrm{RSS}_0/\mathrm{RSS}_1)$, and the heterogeneous case is the same calculation
within each cell. (b) Since $A_w$ is a component of $G'\setminus S$, the global Markov property
gives $X_{A_w}\perp X_{V\setminus(A_w\cup S)}\mid X_S$. As $A_w$ is continuous, the conditional
distribution of $Y_{A_w}$ given $X_S$ is Gaussian with mean linear in $y_{\Gamma_S}$ and
intercept depending on $i_{\Delta_S}$, with covariance common to the cells (homogeneous) or
depending on $i_{\Delta_S}$ (heterogeneous). Integrating out $Y_{A_w\setminus\{w\}}$ gives the
stated regression, which does not involve $z$ and is unchanged by conditioning on all
discrete data and on the continuous variables outside $A_w$. Given these, the design matrices
are fixed and the normal-theory distribution of nested least-squares fits applies: $\mathrm{RSS}_1
\sim\sigma^2\chi^2_{n-r_1}$ independent of $\mathrm{RSS}_0-\mathrm{RSS}_1\sim\sigma^2\chi^2_{r_1-r_0}$,
and the same within each cell. The design has full rank with probability one when $n$ exceeds
the number of its columns, since the continuous regressors have a density given the cells.
\end{proof}

For one edge, Proposition~\ref{prop:onestep} is classical in the homogeneous model: a
decomposable edge-deletion test with a continuous endpoint is an $F$ test
\citep{Frydenberg+Lauritzen:1989}, \citep[Section~5.3]{Edwards:2000}, and the Beta
parameters are derived by \citet{Tur+Castelo:2011} and \citet{Tur+Roverato+Castelo:2014};
\citet[pp.~192--194]{Lauritzen:1996} gives the exact null distributions conditionally on
the counts, in homogeneous and heterogeneous models. We give the proof because
Proposition~\ref{prop:chain} rests on part (b), the formulation conditional on the
discrete data through the continuous side of the graph.
For a continuous edge $\exp(-D/n)=1-r^2$ with $r$ the partial correlation of $Y_u$ and $Y_v$, which
is symmetric in $u$ and $v$, so that the distribution holds whichever endpoint the computation
uses. With $K_S$ occupied cells, $r_0=K_S+g_S$ and $r_1=r_0+1$ for a continuous edge, and
$r_1=K_C+g_S$ for a discrete--continuous edge, which gives the shapes of
Section~\ref{sec:exact}; without discrete variables they reduce to the Gaussian shapes
$((n-|C|)/2,1/2)$.

\begin{proposition}\label{prop:chain}
Let $G_0\subset G_1$ be strongly decomposable, and let every edge of $G_1$ not in $G_0$ have a
continuous endpoint. (a) There are strongly decomposable graphs
$G_1=G^{(0)}\supset G^{(1)}\supset\cdots\supset G^{(J)}=G_0$ differing by one edge, each removed
edge lying in a unique maximal clique of $G^{(i-1)}$ and of its star-augmented graph.
(b) Suppose that the regression design of every step has full column rank and, in the
heterogeneous model, that every group has more observations than columns; these conditions
concern the discrete data only. Then under $G_0$, conditionally on the discrete
observations, the step statistics $B_1,\ldots,B_J$ (homogeneous), or the step deviances
(heterogeneous), are independent.
\end{proposition}

\begin{proof}
(a) This is Lemma~7 of \citet{Frydenberg+Lauritzen:1989}. The star-augmented graphs of
$G_0$ and $G_1$ are chordal and nested, and by their Lemma~5 there is a sequence of chordal
graphs between them, each obtained from the previous one by removing a single edge; every
graph of the sequence contains all star edges and is therefore the star-augmented graph of
a strongly decomposable graph. Removing an edge keeps a graph chordal exactly when the edge
lies in a unique maximal clique (compare their Lemma~3).
(b) Condition on the discrete observations $i^{(1)},\ldots,i^{(n)}$. Under the CG model
$G^{(i)}$ the rows of $Y$ are then independent, row $r$ with density proportional to
$\exp\{h(i^{(r)})^\top y-\tfrac12y^\top K(i^{(r)})y\}$, with $K(i)$ constant in the
homogeneous model. The graph restricts the canonical parameters linearly: $h_\gamma(i)$ is
a sum of interaction terms in the discrete variables of the cliques that contain $\gamma$,
and $K_{\gamma\gamma'}(i)$ vanishes for non-adjacent $\gamma$ and $\gamma'$ and is, in the
heterogeneous model, a sum of interaction terms in the discrete variables of the cliques
that contain both. Only the values of these terms on the occupied cells enter the
likelihood; keeping one parameter for each linearly independent term on the occupied cells
gives a minimal representation, whose canonical statistic $T_i$ collects, for every clique
$C$, the sums of $y_\gamma$, $\gamma\in\Gamma_C$, over the occupied cells of $\Delta_C$
and the sums of $y_\gamma y_{\gamma'}$, $\gamma,\gamma'\in\Gamma_C$, overall (homogeneous)
or within these cells (heterogeneous). The parameter set, with the linear terms free and
the concentration matrices positive definite on every occupied cell, is open in the space
of canonical parameters. The family is therefore a regular exponential family of full rank,
and $T_i$ is complete \citep[Section~4.3]{Lehmann+Romano:2005}. The cell probabilities,
whose estimates may lie on the boundary (Section~\ref{sec:likelihood}), do not enter this
conditional family, so that boundary estimates and empty cells do not affect the argument.
By Proposition~\ref{prop:onestep}(b), under the conditions of (b), $B_i$ has the same
distribution under every member of $G^{(i)}$, that is, it is ancillary. Each later statistic $B_j$, $j>i$, is computed from the
within-cell sums and cross-products of the vertex sets $S_j$, $S_jz$, $S_jw$ and $C_j$, which
are complete in $G^{(j-1)}\subseteq G^{(i)}$, and is therefore a function of $T_i$. By Basu's
theorem \citep{Basu:1955}, $B_i$ is independent of $(B_{i+1},\ldots,B_J)$ under every member
of $G^{(i)}$, in particular under $G_0$; the same argument for $i=1,\ldots,J-1$ gives mutual
independence.
\end{proof}

The factorisation of the likelihood ratio along such a sequence is Proposition~9 of
\citet{Frydenberg+Lauritzen:1989}, who remark that in general the factors are independent
only asymptotically, so that the exact distribution of their product can be complicated.
\citet[Proposition~6.27]{Lauritzen:1996} states that when every removed edge has a
continuous endpoint the factors are exactly independent, conditionally on the counts, and
outlines the proof by Basu's theorem; Proposition~\ref{prop:chain} gives the complete
argument, including the existence of a removal sequence through strongly decomposable
graphs and the conditions on the discrete data under which every step is defined.
Since $\Lambda$ is the difference of the deviances of the two models, it does not depend on the
removal sequence, so that the products obtained from different sequences have the same
distribution. The conditions under which the test applies, $n$ larger than the number of
regression columns of each step and, in the heterogeneous model, enough observations in every
group, depend on the discrete data only, so that leaving out data sets that fail them, as in
Table~\ref{tab:exact-level}, does not affect the exactness conditional on the discrete data.

When edges between two discrete variables are removed, the conditioning is on the margins of
the discrete data instead. Write $I=(i^1,\ldots,i^n)$ for the discrete data and $T(I)$ for the
margins of the cliques of $G_0[\Delta]$, the subgraph of $G_0$ on the discrete vertices.

\begin{proposition}\label{prop:discrete}
Let $G_0\subset G_1$ be strongly decomposable, with independent observations from a
distribution in the conditional Gaussian model of $G_0$ (homogeneous or heterogeneous).
(a) The graph $G_0'=G_0\cup\{\text{removed edges between discrete vertices}\}$ is strongly
decomposable, and every removed edge between discrete vertices lies in a clique of discrete
vertices of the graph from which it is removed. (b) The conditional distribution of $I$ given
$T(I)$ is uniform on the sequences with these margins, whatever the parameters, and the
sampler of Section~\ref{sec:exact} draws from it, up to a permutation of the observations.
(c) Conditionally on $I$, the deviances of the steps from $G_1$ to $G_0'$ are independent, with
the laws of Section~\ref{sec:exact}, and the deviances of the steps from $G_0'$ to $G_0$ are
functions of $I$. Hence the conditional distribution of $\Lambda$ given $T(I)$ is free of the
parameters, and the simulated p-value estimates an exact conditional p-value.
\end{proposition}

\begin{proof}
(a) The existence of a sequence from $G_0$ to $G_1$ through strongly decomposable graphs in
which the edges between discrete vertices are added first is Lemma~2.22 of
\citet{Lauritzen:1996}; its graph after those additions is $G_0'$. If a removed edge
$\{u,v\}$ between discrete vertices lay in a clique with a continuous vertex $w$, the path
$u,w,v$ would be a forbidden path of the graph after the removal.
(b) In a strongly decomposable graph the boundary of every connected component of the
continuous vertices is a complete set of discrete vertices, since two non-adjacent discrete
vertices of that boundary would be joined by a forbidden path. The model of $G_0$ is therefore
collapsible onto $\Delta$ \citep{Frydenberg:1990}: the density of an observation is
$p(i)f(y\mid i)$, where $p$ ranges over the decomposable log-linear model of $G_0[\Delta]$ and
the parameters of $f(y\mid i)$ are variation independent of $p$. The likelihood of $I$ is
$\prod_\nu p(i^\nu)$, a function of the parameters and of $T(I)$ only, so that, given $T(I)$,
every sequence with these margins has the same probability. The sampler keeps the first clique
and, for each later clique $C_j$ with separator $S_j$ and $R_j=C_j\setminus S_j$, assigns the
observed values of $R_j$ within each cell of $S_j$ to the observations of that cell in random
order. The number of distinct assignments at step $j$ depends only on the margins of $C_j$, so
that every completion is equally likely and the result is uniform on the sequences with the
margins $T(I)$ and the observed first clique. $\Lambda$ and the conditional laws in (c) depend on
$I$ only through its table of counts, which is invariant under permutations of the
observations, and every sequence with the margins $T(I)$ is a permutation of one with the
observed first clique.
(c) Under $G_0$ the data follow the model of $G_0'\supseteq G_0$, and every edge of $G_1$ not in
$G_0'$ has a continuous endpoint, so Proposition~\ref{prop:chain} applies to $G_0'\subset G_1$.
The steps from $G_0'$ to $G_0$ are contingency-table deviances in cliques of discrete vertices,
by (a). The joint conditional law of $I$ and the step deviances given $T(I)$ is thus the law
of (b) followed by the laws of Proposition~\ref{prop:chain} given $I$, free of the parameters.
Leaving out simulated data sets on which a continuous step is undefined conditions further on
an event of $I$, which preserves exactness.
\end{proof}

This is the procedure that \citet[p.~196]{Lauritzen:1996} outlines for evaluating the tail
probability of a joint test; in purely discrete models it is the exact conditional test of
nested decomposable log-linear models.


\newpage

\begin{thebibliography}{99}

\bibitem[Abreu et al.(2010)]{Abreu+Edwards+Labouriau:2010}
Abreu, G.~C.~G., Edwards, D. and Labouriau, R. (2010).
High-dimensional graphical model search with the \pkg{gRapHD} \proglang{R} package.
\textit{Journal of Statistical Software} \textbf{37}(1), 1--18.
doi:10.18637/jss.v037.i01.

\bibitem[Aeberhard and Forina(1992)]{Aeberhard+Forina:1992}
Aeberhard, S. and Forina, M. (1992).
Wine [data set].
UCI Machine Learning Repository.
doi:10.24432/C5PC7J.

\bibitem[Akaike(1974)]{Akaike:1974}
Akaike, H. (1974).
A new look at the statistical model identification.
\textit{IEEE Transactions on Automatic Control} \textbf{19}(6), 716--723.
doi:10.1109/TAC.1974.1100705.

\bibitem[Anderson(2003)]{Anderson:2003}
Anderson, T.~W. (2003).
\textit{An Introduction to Multivariate Statistical Analysis},
3rd ed.
Wiley-Interscience, Hoboken, NJ.

\bibitem[Basu(1955)]{Basu:1955}
Basu, D. (1955).
On statistics independent of a complete sufficient statistic.
\textit{Sankhy\={a}} \textbf{15}(4), 377--380.

\bibitem[Beerenwinkel et al.(2007)]{Beerenwinkel+Eriksson+Sturmfels:2007}
Beerenwinkel, N., Eriksson, N. and Sturmfels, B. (2007).
Conjunctive Bayesian networks.
\textit{Bernoulli} \textbf{13}(4), 893--909.
doi:10.3150/07-BEJ6133.

\bibitem[Chen and Chen(2008)]{Chen+Chen:2008}
Chen, J. and Chen, Z. (2008).
Extended Bayesian information criteria for model selection with large model
spaces.
\textit{Biometrika} \textbf{95}(3), 759--771.
doi:10.1093/biomet/asn034.

\bibitem[Chow and Liu(1968)]{Chow+Liu:1968}
Chow, C.~K. and Liu, C.~N. (1968).
Approximating discrete probability distributions with dependence trees.
\textit{IEEE Transactions on Information Theory} \textbf{14}(3), 462--467.
doi:10.1109/TIT.1968.1054142.

\bibitem[Davidson and MacKinnon(2007)]{Davidson+MacKinnon:2007}
Davidson, R. and MacKinnon, J.~G. (2007).
Improving the reliability of bootstrap tests with the fast double bootstrap.
\textit{Computational Statistics \& Data Analysis} \textbf{51}(7), 3259--3281.
doi:10.1016/j.csda.2006.04.001.

\bibitem[Davison and Hinkley(1997)]{Davison+Hinkley:1997}
Davison, A.~C. and Hinkley, D.~V. (1997).
\textit{Bootstrap Methods and their Application}.
Cambridge University Press, Cambridge.

\bibitem[Dempster et al.(1977)]{Dempster+Laird+Rubin:1977}
Dempster, A.~P., Laird, N.~M. and Rubin, D.~B. (1977).
Maximum likelihood from incomplete data via the EM algorithm.
\textit{Journal of the Royal Statistical Society, Series B (Methodological)}
\textbf{39}(1), 1--38.
doi:10.1111/j.2517-6161.1977.tb01600.x.

\bibitem[Desper et al.(1999)]{Desper+others:1999}
Desper, R., Jiang, F., Kallioniemi, O.-P., Moch, H., Papadimitriou, C.~H. and
Sch\"affer, A.~A. (1999).
Inferring tree models for oncogenesis from comparative genome hybridization data.
\textit{Journal of Computational Biology} \textbf{6}(1), 37--51.
doi:10.1089/cmb.1999.6.37.

\bibitem[Dethlefsen and H{\o}jsgaard(2005)]{Dethlefsen+Hojsgaard:2005}
Dethlefsen, C. and H{\o}jsgaard, S. (2005).
A common platform for graphical models in \proglang{R}: The \pkg{gRbase} package.
\textit{Journal of Statistical Software} \textbf{14}(17), 1--12.
doi:10.18637/jss.v014.i17.

\bibitem[Detrano et al.(1989)]{Detrano+others:1989}
Detrano, R., Janosi, A., Steinbrunn, W., Pfisterer, M., Schmid, J.-J., Sandhu, S., Guppy,
K.~H., Lee, S. and Froelicher, V. (1989).
International application of a new probability algorithm for the diagnosis of coronary
artery disease.
\textit{The American Journal of Cardiology} \textbf{64}(5), 304--310.
doi:10.1016/0002-9149(89)90524-9.

\bibitem[Drton and Plummer(2017)]{Drton+Plummer:2017}
Drton, M. and Plummer, M. (2017).
A Bayesian information criterion for singular models.
\textit{Journal of the Royal Statistical Society, Series B (Statistical Methodology)}
\textbf{79}(2), 323--380.
doi:10.1111/rssb.12187.

\bibitem[Edwards(2000)]{Edwards:2000}
Edwards, D. (2000).
\textit{Introduction to Graphical Modelling}, 2nd ed.
Springer-Verlag, New York.

\bibitem[Edwards et al.(2010)]{Edwards+Abreu+Labouriau:2010}
Edwards, D., de Abreu, G.~C.~G. and Labouriau, R. (2010).
Selecting high-dimensional mixed graphical models using minimal AIC or BIC forests.
\textit{BMC Bioinformatics} \textbf{11}, 18.
doi:10.1186/1471-2105-11-18.

\bibitem[Efron(1994)]{Efron:1994}
Efron, B. (1994).
Missing data, imputation, and the bootstrap.
\textit{Journal of the American Statistical Association}
\textbf{89}(426), 463--475.
doi:10.1080/01621459.1994.10476768.

\bibitem[Fienberg and Rinaldo(2012)]{Fienberg+Rinaldo:2012}
Fienberg, S.~E. and Rinaldo, A. (2012).
Maximum likelihood estimation in log-linear models.
\textit{The Annals of Statistics} \textbf{40}(2), 996--1023.
doi:10.1214/12-AOS986.

\bibitem[Fiorini(2016)]{Fiorini:2016}
Fiorini, S. (2016).
Gene expression cancer RNA-Seq [data set].
UCI Machine Learning Repository.
doi:10.24432/C5R88H.

\bibitem[Foygel and Drton(2010)]{Foygel+Drton:2010}
Foygel, R. and Drton, M. (2010).
Extended Bayesian information criteria for Gaussian graphical models.
In \textit{Advances in Neural Information Processing Systems 23},
pp.~604--612.
arXiv:1011.6640.

\bibitem[Friedman(1997)]{Friedman:1997}
Friedman, N. (1997).
Learning belief networks in the presence of missing values and hidden variables.
In D.~H. Fisher (ed.),
\textit{Proceedings of the Fourteenth International Conference on Machine Learning
(ICML 1997)}, pp.~125--133.
Morgan Kaufmann, San Francisco.

\bibitem[Friedman(1998)]{Friedman:1998}
Friedman, N. (1998).
The Bayesian structural EM algorithm.
In G.~F. Cooper and S. Moral (eds.),
\textit{Proceedings of the Fourteenth Conference on Uncertainty in Artificial
Intelligence (UAI '98)}, pp.~129--138.
Morgan Kaufmann, San Francisco.

\bibitem[Friedman et al.(2008)]{Friedman+Hastie+Tibshirani:2008}
Friedman, J., Hastie, T. and Tibshirani, R. (2008).
Sparse inverse covariance estimation with the graphical lasso.
\textit{Biostatistics} \textbf{9}(3), 432--441.
doi:10.1093/biostatistics/kxm045.

\bibitem[Frydenberg(1990)]{Frydenberg:1990}
Frydenberg, M. (1990).
Marginalization and collapsibility in graphical interaction models.
\textit{The Annals of Statistics} \textbf{18}(2), 790--805.
doi:10.1214/aos/1176347626.

\bibitem[Frydenberg and Lauritzen(1989)]{Frydenberg+Lauritzen:1989}
Frydenberg, M. and Lauritzen, S.~L. (1989).
Decomposition of maximum likelihood in mixed graphical interaction models.
\textit{Biometrika} \textbf{76}(3), 539--555.
doi:10.1093/biomet/76.3.539.

\bibitem[Gerstung et al.(2009)]{Gerstung+others:2009}
Gerstung, M., Baudis, M., Moch, H. and Beerenwinkel, N. (2009).
Quantifying cancer progression with conjunctive Bayesian networks.
\textit{Bioinformatics} \textbf{25}(21), 2809--2815.
doi:10.1093/bioinformatics/btp505.

\bibitem[Giudici and Green(1999)]{Giudici+Green:1999}
Giudici, P. and Green, P.~J. (1999).
Decomposable graphical Gaussian model determination.
\textit{Biometrika} \textbf{86}(4), 785--801.
doi:10.1093/biomet/86.4.785.

\bibitem[Gusfield(1991)]{Gusfield:1991}
Gusfield, D. (1991).
Efficient algorithms for inferring evolutionary trees.
\textit{Networks} \textbf{21}(1), 19--28.
doi:10.1002/net.3230210104.

\bibitem[H{\o}jsgaard et al.(2012)]{Hojsgaard+Edwards+Lauritzen:2012}
H{\o}jsgaard, S., Edwards, D. and Lauritzen, S. (2012).
\textit{Graphical Models with \proglang{R}}.
Springer, New York.
doi:10.1007/978-1-4614-2299-0.

\bibitem[Honaker and King(2010)]{Honaker+King:2010}
Honaker, J. and King, G. (2010).
What to do about missing values in time-series cross-section data.
\textit{American Journal of Political Science}
\textbf{54}(2), 561--581.
doi:10.1111/j.1540-5907.2010.00447.x.

\bibitem[Janosi et al.(1989)]{Janosi+others:1989}
Janosi, A., Steinbrunn, W., Pfisterer, M. and Detrano, R. (1989).
Heart disease [data set].
UCI Machine Learning Repository.
doi:10.24432/C52P4X.

\bibitem[Julier and Uhlmann(2004)]{Julier+Uhlmann:2004}
Julier, S.~J. and Uhlmann, J.~K. (2004).
Unscented filtering and nonlinear estimation.
\textit{Proceedings of the IEEE} \textbf{92}(3), 401--422.
doi:10.1109/JPROC.2003.823141.

\bibitem[Kirkpatrick et al.(1983)]{Kirkpatrick+Gelatt+Vecchi:1983}
Kirkpatrick, S., Gelatt, C.~D., Jr., and Vecchi, M.~P. (1983).
Optimization by simulated annealing.
\textit{Science} \textbf{220}(4598), 671--680.
doi:10.1126/science.220.4598.671.

\bibitem[Kruskal(1956)]{Kruskal:1956}
Kruskal, J.~B. (1956).
On the shortest spanning subtree of a graph and the traveling salesman problem.
\textit{Proceedings of the American Mathematical Society} \textbf{7}(1), 48--50.
doi:10.1090/S0002-9939-1956-0078686-7.

\bibitem[Labouriau(2026)]{Labouriau:AccumulationTrees}
Labouriau, R. (2026).
Graphical models induced by irreversible event accumulation on trees.
arXiv:2610.40127 [math.ST].
doi:10.48550/arXiv.2610.40127.

\bibitem[Lauritzen(1995)]{Lauritzen:1995}
Lauritzen, S.~L. (1995).
The EM algorithm for graphical association models with missing data.
\textit{Computational Statistics \& Data Analysis} \textbf{19}(2), 191--201.
doi:10.1016/0167-9473(93)E0056-A.

\bibitem[Lauritzen(1996)]{Lauritzen:1996}
Lauritzen, S.~L. (1996).
\textit{Graphical Models}.
Oxford University Press, Oxford.

\bibitem[Lauritzen and Wermuth(1989)]{Lauritzen+Wermuth:1989}
Lauritzen, S.~L. and Wermuth, N. (1989).
Graphical models for associations between variables, some of which are
qualitative and some quantitative.
\textit{The Annals of Statistics} \textbf{17}(1), 31--57.
doi:10.1214/aos/1176347003.

\bibitem[Lehmann and Romano(2005)]{Lehmann+Romano:2005}
Lehmann, E.~L. and Romano, J.~P. (2005).
\textit{Testing Statistical Hypotheses}, 3rd ed.
Springer, New York.
doi:10.1007/0-387-27605-X.

\bibitem[Little and Rubin(2019)]{Little+Rubin:2019}
Little, R.~J.~A. and Rubin, D.~B. (2019).
\textit{Statistical Analysis with Missing Data},
3rd ed.
John Wiley \& Sons, Hoboken, NJ.
doi:10.1002/9781119482260.

\bibitem[Lourenço et al.(2003)]{Lourenco+Martin+Stutzle:2003}
Lourenço, H.~R., Martin, O.~C., and Stützle, T. (2003).
Iterated local search.
In F.~Glover and G.~A. Kochenberger (eds.),
\textit{Handbook of Metaheuristics}, pp.~320--353.
Kluwer Academic Publishers, Boston.
doi:10.1007/0-306-48056-5\_11.

\bibitem[Patefield(1981)]{Patefield:1981}
Patefield, W.~M. (1981).
Algorithm AS 159: An efficient method of generating random $r\times c$ tables with given
row and column totals.
\textit{Journal of the Royal Statistical Society, Series C (Applied Statistics)}
\textbf{30}(1), 91--97.
doi:10.2307/2346669.

\bibitem[Rubin(1987)]{Rubin:1987}
Rubin, D.~B. (1987).
\textit{Multiple Imputation for Nonresponse in Surveys}.
John Wiley \& Sons, New York.
doi:10.1002/9780470316696.

\bibitem[Schwarz(1978)]{Schwarz:1978}
Schwarz, G. (1978).
Estimating the dimension of a model.
\textit{The Annals of Statistics} \textbf{6}(2), 461--464.
doi:10.1214/aos/1176344136.

\bibitem[Szabo and Boucher(2002)]{Szabo+Boucher:2002}
Szabo, A. and Boucher, K. (2002).
Estimating an oncogenetic tree when false negatives and positives are present.
\textit{Mathematical Biosciences} \textbf{176}(2), 219--236.
doi:10.1016/S0025-5564(02)00086-X.

\bibitem[Tur and Castelo(2011)]{Tur+Castelo:2011}
Tur, I. and Castelo, R. (2011).
Learning mixed graphical models from data with $p$ larger than $n$.
In \textit{Proceedings of the 27th Conference on Uncertainty in Artificial Intelligence
(UAI 2011)}, pp.~689--697. AUAI Press, Corvallis, OR.
arXiv:1202.3765.

\bibitem[Tur et al.(2014)]{Tur+Roverato+Castelo:2014}
Tur, I., Roverato, A. and Castelo, R. (2014).
Mapping eQTL networks with mixed graphical Markov models.
\textit{Genetics} \textbf{198}(4), 1377--1393.
doi:10.1534/genetics.114.169573.

\bibitem[U.S. Bureau of the Census(1977)]{USCensus:1977}
U.S. Department of Commerce, Bureau of the Census (1977).
\textit{Statistical Abstract of the United States}, and
\textit{County and City Data Book}.
Washington, DC. Distributed with \proglang{R} as \code{state.x77} in the package
\pkg{datasets}.

\bibitem[Weinstein et al.(2013)]{Weinstein+others:2013}
Weinstein, J.~N., Collisson, E.~A., Mills, G.~B., Shaw, K.~R.~M., Ozenberger, B.~A.,
Ellrott, K., Shmulevich, I., Sander, C., Stuart, J.~M. and The Cancer Genome Atlas
Research Network (2013).
The Cancer Genome Atlas Pan-Cancer analysis project.
\textit{Nature Genetics}, \textbf{45}(10), 1113--1120.
doi:10.1038/ng.2764.

\bibitem[Whittaker(1990)]{Whittaker:1990}
Whittaker, J. (1990).
\textit{Graphical Models in Applied Multivariate Statistics}.
John Wiley \& Sons, Chichester.

\end{thebibliography}
\end{document}